\documentclass[linenumbers, twocolumn]{aastex631}
\hypersetup{linkcolor=blue,citecolor=blue,filecolor=cyan,urlcolor=blue}

\usepackage{tabularx}
\usepackage{url}
\usepackage{graphicx}
\usepackage[T1]{fontenc}
\usepackage{ae,aecompl}
\usepackage{booktabs}
\usepackage{natbib}
\usepackage{enumitem}
\usepackage{mathtools}
\usepackage{amssymb}
\usepackage{amsmath}
\usepackage{ulem}
\usepackage{epstopdf}
\usepackage{color}
\usepackage{xcolor}
\usepackage{needspace}
\usepackage{soul}

\usepackage{blindtext}
\usepackage{longtable}
\usepackage{tabu}
\usepackage{lineno}

\def\unit #1{\,{\rm #1}}

\newcommand\cmsqi{\rm \,\unit{cm^{-2}}}

\newcommand\kev{\rm \,\unit{keV}}

\newcommand\funit{\rm \,erg\,cm^{-2}\,s^{-1}}

\newcommand\lunit{\rm \,erg \,s^{-1}}

\newcommand\ledd{L_{\rm Edd}}

\newcommand\lambdaedd{\lambda_{\rm Edd}}

\newcommand\lbol{L_{\rm  bol}}

\newcommand\funita{\rm\,erg\,cm^{-2}\,s^{-1}\angstrom^{-1}}

\newcommand\msol{M_{\odot}}

\newcommand\nh{\rm N_{H}}

\newcommand\pc{\unit{pc}}

\newcommand\ev{\unit{\, eV}}
\usepackage{amsmath} 
\newcommand{\angstrom}{\text{\normalfont\AA}}

\newcommand\swift{{\it Swift}}
\newcommand\xmm{{\it XMM-Newton}}

\newcommand\nustar{{\it NuSTAR}}

\def\aap{A\&A} \def\apjl{ApJ}  
 \def\apjs{ApJS}

\shorttitle{1ES1927+654 multi-wavelength view of jet formation and evolution}

\begin{document}
\nolinenumbers



\title{Multi-wavelength observations of a jet launch in real time from the  post-changing-look Active Galaxy 1ES 1927+654}

\author[0000-0003-2714-0487]{Sibasish Laha} 
\affiliation{Astrophysics Science Division, NASA Goddard Space Flight Center,Greenbelt, MD 20771, USA.}
\affiliation{Center for Space Science and Technology, University of Maryland Baltimore County, 1000 Hilltop Circle, Baltimore, MD 21250, USA.}
\affiliation{Center for Research and Exploration in Space Science and Technology, NASA/GSFC, Greenbelt, Maryland 20771, USA}

\author[0000-0002-7676-9962]{Eileen T. Meyer}
\affiliation{Department of Physics, University of Maryland Baltimore County, 1000 Hilltop Circle Baltimore, MD 21250, USA}

\author[0000-0003-2714-0487]{Dev R. Sadaula} 
\affiliation{Astrophysics Science Division, NASA Goddard Space Flight Center, Greenbelt, MD 20771, USA.}
\affiliation{Center for Space Science and Technology, University of Maryland Baltimore County, 1000 Hilltop Circle, Baltimore, MD 21250, USA.}
\affiliation{Center for Research and Exploration in Space Science and Technology, NASA/GSFC, Greenbelt, Maryland 20771, USA}

\author[0000-0003-4790-2653]{Ritesh Ghosh}
\affiliation{Center for Space Science and Technology, University of Maryland Baltimore County, 1000 Hilltop Circle, Baltimore, MD 21250, USA.}
\affiliation{Astrophysics Science Division, NASA Goddard Space Flight Center, Greenbelt, MD 20771, USA.}
\affiliation{Center for Research and Exploration in Space Science and Technology, NASA/GSFC, Greenbelt, Maryland 20771, USA}
\affiliation{MKHS, Murshidabad, West Bengal, India 742401, India }

\author[0009-0002-6991-1534]{Dhrubojyoti Sengupta}
\affiliation{Center for Space Science and Technology, University of Maryland Baltimore County, 1000 Hilltop Circle, Baltimore, MD 21250, USA.}

\author[0000-0003-4127-0739]{Megan Masterson}
\affiliation{MIT Kavli Institute for Astrophysics and Space Research, Massachusetts Institute of Technology, Cambridge, MA 02139, USA}

\author[0000-0003-4727-2209]{Onic I. Shuvo}
\affiliation{Department of Physics, University of Maryland Baltimore County, 1000 Hilltop Circle Baltimore, MD 21250, USA}

\author[0000-0002-1094-3147]{Matteo Guainazzi}
\affiliation{European Space Agency (ESA), European Space Research and Technology Centre (ESTEC), Keplerlaan 1, 2201 AZ Noordwijk, The Netherlands}

\author[0000-0002-1094-3147]{Claudio Ricci}
\affiliation{Instituto de Estudios Astrof\'isicos, Facultad de Ingenier\'ia y Ciencias, Universidad Diego Portales, Av. Ej\'ercito Libertador 441, Santiago, Chile}
\affiliation{Kavli Institute for Astronomy and Astrophysics, Peking University, Beijing 100871, China}


\author[0000-0003-0936-8488]{Mitchell C.~Begelman}
\affiliation{JILA, University of Colorado and National Institute of Standards and Technology, 440 UCB, Boulder, CO 80309-0440, USA.}
\author[0000-0001-7801-0362]{Alexander Philippov}
\affiliation{Department of Physics, University of Maryland, College Park, MD, USA}
\author[0000-0001-9475-5292]{Rostom Mbarek}
\altaffiliation{Neil Gehrels Fellow}
\affiliation{Joint Space-Science Institute, University of Maryland, College Park, MD, USA}
\affiliation{Department of Astronomy, University of Maryland, College Park, MD, USA}
\author[0000-0001-9725-5509]{Amelia M. Hankla}
\altaffiliation{Neil Gehrels Fellow}
\affiliation{Joint Space-Science Institute, University of Maryland, College Park, MD, USA}
\affiliation{Department of Astronomy, University of Maryland, College Park, MD, USA}


\author[0000-0003-0172-0854]{Erin Kara}
\affiliation{MIT Kavli Institute for Astrophysics and Space Research, Massachusetts Institute of Technology, Cambridge, MA 02139, USA}


\author[0000-0003-0543-3617]{Francesca Panessa}
\affiliation{INAF -- Istituto di Astrofisica e Planetologia Spaziali, Via del Fosso del Cavaliere 100, Roma, 00133, Italy}

\author[0000-0000-0000-0000]{Ehud Behar}
\affiliation{Department of Physics, Technion, Haifa 32000, Israel}



\author[0000-0000-0000-0000]{Haocheng Zhang}
\affiliation{Astrophysics Science Division, NASA Goddard Space Flight Center,Greenbelt, MD 20771, USA.}
\affiliation{Center for Space Science and Technology, University of Maryland Baltimore County, 1000 Hilltop Circle, Baltimore, MD 21250, USA.}
\affiliation{Center for Research and Exploration in Space Science and Technology, NASA/GSFC, Greenbelt, Maryland 20771, USA}

\author[0000-0001-9879-7780]{Fabio Pacucci}
\affiliation{Center for Astrophysics $\vert$ Harvard \& Smithsonian, Cambridge, MA 02138, USA} 
\affiliation{Black Hole Initiative, Harvard University, Cambridge, MA 02138, USA}


\author[0000-0001-6523-6522]{Main Pal}
\affiliation{Department of Physics, Sri Venkateswara College, University of Delhi, Benito Juarez Road, Dhaula Kuan,  New Delhi -- 110021, India}

\author[0000-0001-5742-5980]{Federica Ricci}
\affiliation{Dipartimento di Matematica e Fisica, Universit\`a degli Studi Roma Tre, Via della Vasca Navale 84, 00146, Roma, Italy}
\affiliation{INAF-Osservatorio Astronomico di Roma, via Frascati 33, 00040 Monteporzio Catone, Italy}

\author[0009-0006-7483-0463]{Ilaria Villani}
\affiliation{Dipartimento di Matematica e Fisica, Universit\`a degli Studi Roma Tre, Via della Vasca Navale 84, 00146, Roma, Italy}
\affiliation{INAF-Osservatorio Astronomico di Roma, via Frascati 33, 00040 Monteporzio Catone, Italy}

\author[0000-0003-3746-4565]{Susanna Bisogni}
\affiliation{INAF – Istituto di Astrofisica Spaziale e Fisica Cosmica Milano, Via Corti 12, 20133 Milano, Italy}

\author[0000-0000-0000-0000]{Fabio La Franca}
\affiliation{Dipartimento di Matematica e Fisica, Universit\`a degli Studi Roma Tre, Via della Vasca Navale 84, 00146, Roma, Italy}

\author[0000-0002-4622-4240]{Stefano Bianchi}
\affiliation{Dipartimento di Matematica e Fisica, Universit\`a degli Studi Roma Tre, Via della Vasca Navale 84, 00146, Roma, Italy}

\author[0000-0002-5182-6289]{Gabriele Bruni}
\affiliation{INAF -- Istituto di Astrofisica e Planetologia Spaziali, Via del Fosso del Cavaliere 100, Roma, 00133, Italy}

\author[0000-0000-0000-0000]{Samantha Oates}

\affiliation{Department of Physics, Lancaster University, Lancaster LA1 4YB, UK}

\author[0000-0000-0000-0000]{Cameron Hahn}
\affiliation{Department of Physics, University of Maryland Baltimore County, 1000 Hilltop Circle Baltimore, MD 21250, USA}
\author[0000-0002-2555-3192]{Matt Nicholl}

\affiliation{Astrophysics Research Centre, School of Mathematics and Physics, Queens University Belfast, Belfast BT7 1NN, UK}

\author[0000-0003-1673-970X]{S. Bradley Cenko}
\affiliation{Astrophysics Science Division, NASA Goddard Space Flight Center,Greenbelt, MD 20771, USA.}
\affiliation{Joint Space-Science Institute, University of Maryland, College Park, MD 20742, USA}

\author[0000-0003-1601-8048]{Sabyasachi Chattopadhyay}
\affiliation{South African Astronomical Observatory, 1 Observatory Rd, Observatory,
Cape Town, 7925, South Africa}
\affiliation{Centre for Space Research, North-West University, Potchefstroom 2520,
South Africa}


\author[0000-0000-0000-0000]{Josefa Becerra  Gonz\'alez} 
\affiliation{Instituto de Astrof\'isica de Canarias (IAC), E-38200 La Laguna, Tenerife, Spain}
\affiliation{Universidad de La Laguna (ULL), Departamento de Astrof\'isica, E-38206 La Laguna, Tenerife, Spain}

\author[0000-0000-0000-0000]{J.~A.~Acosta--Pulido}
\affiliation{Instituto de Astrof\'isica de Canarias (IAC), E-38200 La Laguna, Tenerife, Spain}
\affiliation{Universidad de La Laguna (ULL), Departamento de Astrof\'isica, E-38206 La Laguna, Tenerife, Spain}

\author[0000-0002-8377-9667]{Suvendu Rakshit}
\affiliation{Aryabhatta Research Institute of Observational Sciences (ARIES), Manora Peak, Nainital, 263002 India}

\author[0000-0003-2931-0742]{Ji\v{r}\'{i} Svoboda\thanks{E-mail: jiri.svoboda@asu.cas.cz}}
\affiliation{Astronomical Institute of the Czech Academy of Sciences, Bo\v{c}n\'{i} II 1401/1, 14100 Praha 4, Czech Republic}

\author[0009-0006-4968-7108]{Luigi Gallo}
\affiliation{Department of Astronomy and Physics, Saint Mary's University, 923 Robie Street, Halifax, B3H 3C3, Canada}

\author[0000-0002-5311-9078]{Adam Ingram}
\affiliation{School of Mathematics Statistics and Physics, Newcastle University, UK, NE1 7RU}

\author[0000-0002-2603-2639]{Darshan Kakkad}
\affiliation{Space Telescope Science Institute, 3700 San Martin Drive, Baltimore, 21210 MD, USA}


\begin{abstract}
\nolinenumbers

 We present results from a high cadence multi-wavelength observational campaign of the enigmatic changing look AGN 1ES 1927+654 from May 2022-
April 2024, coincident with an unprecedented radio flare (an increase in flux by a factor of $\sim 60$ over a few months) and the emergence of a spatially resolved jet at
$0.1-0.3\pc$ scales (Meyer et al. 2024). Companion work has also
detected a recurrent quasi-periodic oscillation (QPO) in the $2-10\kev$
band with an increasing frequency ($1-2$ mHz) over the same period
(Masterson et al., 2025). During this time, the soft X-rays
($0.3-2\kev$) monotonically increased by a factor of  $\sim 8$, while the UV emission remained near-steady with $<30\%$ variation and the
$2-10\kev$ flux showed variation by a factor $\lesssim 2$. The weak
variation of the $2-10\kev$ X-ray emission and the stability of the UV
emission suggest that the magnetic energy density and accretion rate are
relatively unchanged, and that the jet could be launched due to a
reconfiguration of the magnetic field (toroidal to poloidal) close to
the black hole. Advecting poloidal flux onto the event horizon would
trigger the Blandford-Znajek (BZ) mechanism, leading to the onset of the
jet. The concurrent softening of the coronal slope (from $\Gamma=
2.70\pm 0.04$ to $\Gamma=3.27\pm 0.04$), the appearance of a QPO, and low coronal temperature
($kT_{e}=8_{-3}^{+8}\kev$) during the radio outburst suggest that the poloidal field reconfiguration can significantly impact coronal properties and
thus influence jet dynamics. These extraordinary findings in real
time are crucial for coronal and jet plasma studies, particularly as
our results are independent of coronal geometry.

\end{abstract}




\section{Introduction} \label{sec:intro}



Active Galactic Nuclei (AGN) are a diverse class of sources which can be detected at nearly every waveband from radio to gamma-rays, and over a very wide range of luminosity, with bolometric power up to $10^{48}$ erg s$^{-1}$ \citep{duras2020,saccheo2023}. Powered by accretion onto the central super-massive black hole, AGN activity has been shown to impact not only the evolution of the host galaxy but the surrounding environment \citep{kormendy2013}. In part due  to its very small size, the exact physics and geometry of the central engine in AGN is still not clearly understood. For example, we still do not have a clear understanding of the angular momentum transfer mechanisms which help the accretion flow in the disk. The nature of the accretion flow (i.e., accretion mode) varies widely over the $\sim$6-7 orders of magnitude in observed mass accretion rate. Both the accretion-mode and accretion-rate play an important role in the production of jets and outflows which go on to influence not just the immediate vicinity of the central engine and the host galaxy but also the larger-scale environments \citep{fabian2012,begelman2022}.   

Although spectral and flux variability is quite common in AGN across a range of wavelengths \citep{elvis1994, hovatta_radio_variability_2007, ricci2017, laha_coronareview_arxiv_2024}, in some rare cases we find rapid changes in the optical spectral type (type-II to type-I) and/or order of magnitude flares in multi-wavelength bands in a time span of months to years, possibly caused/triggered by rapid accretion-mode and/or accretion-rate changes. This class of AGN is popularly known as `changing-look' AGN \citep{Lamassa2015,trak19,ricci_NatAs_CLAGNreview_2023} and they serve as ideal sources to probe the physics of the central engine in real-time, and its connection with the immediate surroundings \citep{ricci_NatAs_CLAGNreview_2023}.

\begin{figure*}[]
 \centering
{\includegraphics[width=18cm, height=8.5cm]{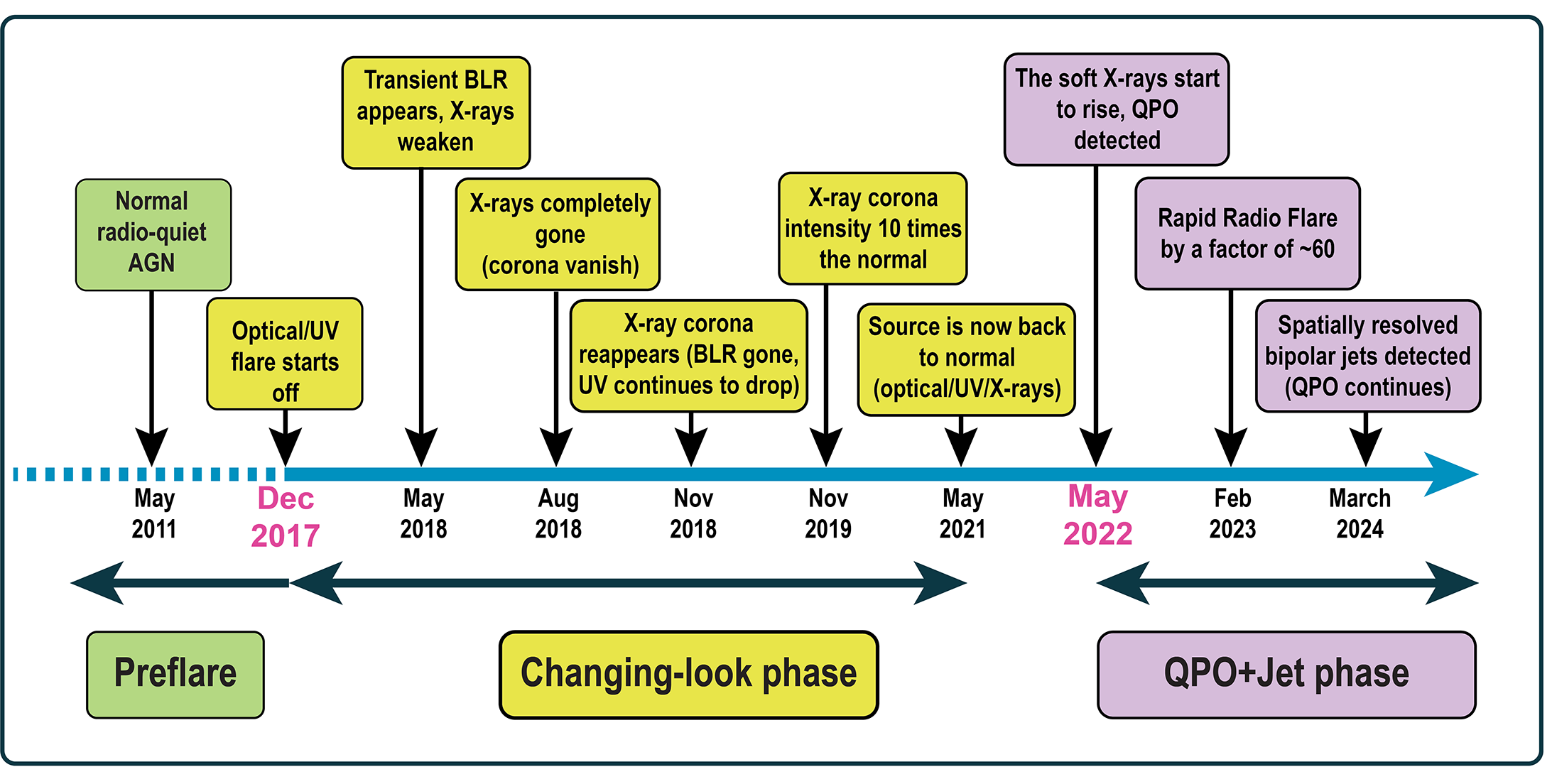}
\caption{The timeline of the enigmatic behavior of the source 1ES~1927+654. From the left to right are the three phases: The preflare phase (in green), the Changing-look phase (in Yellow) and the QPO+Jet phase (in pink). The source exhibited a typical radio-quiet Seyfert galaxy properties in radio, optical/UV and X-rays till Dec 2017 when there was a sudden flare in the optical/UV band which reached a peak of $\sim 100$ times in $\sim 3$ months \citep{trak19}. A transient broad line region (BLR) also appeared during this time, while the X-rays start to weaken by May 2018. This source never exhibited a BLR and was earlier classified as a true type-II source, meaning that there was no line of sight obscuration in optical/UV/X-rays, yet there was no sign of a BLR emission. The X-ray Coronal emission completely vanished in Aug 2018 \citep{Ricci_2020,Ricci2021,laha2022,masterson2022}, and it came back in Oct-Nov 2018. By this time the BLR emission lines (broad Balmer lines) have disappeared, and the UV/Optical flux continues to drop with a $t^{-0.91}$ power law decay. The X-ray corona revived and jumped to $\sim 10$ times that of preflare state in Nov 2019 while the UV continued to drop. The source finally came back to its preflare state in May 2021. Since May 2022 we detected a gradual rise in soft X-rays as well as a quasi-periodic-oscillation (Masterson et al. 2025, Nature, in press). Since Feb 2023 we also detected a rapid radio flare, with a radio flux increase of $\sim 60$ times over a few months. The radio flux continues to stay in the high state for over the next year. Since March 2024 we detected a spatially resolved bipolar jet in the K band, $\sim 22$ GHz \citep{meyer2024}.}
\label{fig:timeline}}
\end{figure*}

1ES 1927+654 is an enigmatic CL-AGN with a supermassive black hole (SMBH) mass of $1.36\times 10^6\msol$ \citep{Li_broad_line_1es} at a redshift of $z=0.017$.  The source first exhibited a dramatic optical/UV outburst starting in December 2017 \citep{trak19,Ricci_2020} and has since showed unprecedented behavior, particularly in the X-rays, when compared to other known CL-AGNs \citep{Ricci_2020,Ricci2021,laha2022,masterson2022}. Earlier optical and X-ray studies \citep{boller2003, gallo2013} have shown that the source was a true type II AGN  pre-outburst \citep{panessa02, bianchi12}: despite having only narrow emission lines in the optical spectrum, there was no evidence of line-of-sight obscuration at any wavelength. After the optical/UV flare in 2018, broad emission lines appeared with a lag of $\sim100$ days and remained visible for several months after.
Conversely, the 2–10 keV coronal hard X-ray emission
completely vanished by 2018 August \citep{Ricci_2020,Ricci2021,laha2022,ricci_NatAs_CLAGNreview_2023} only to return a few months later, eventually flaring up to $\sim 10$ times the pre-flare value. There has been essentially no correlation between the X-rays and the optical/UV both during the 2018 flare and in the several years since. Figure \ref{fig:timeline} illustrates the timeline of the most important events and phases of the source since May 2011.

After the flare, the UV flux monotonically
dimmed with a slope $t^{-0.91\pm 0.04}$ \citep{laha2022}, returning to a near preflare value after $\sim 1200$ days of the initial flare. The parsec-scale radio flux density at 5 GHz, as measured by the Very Long Baseline Array, showed a minimum (a factor of 4 below the preflare value) at the time when the X-ray flux was low, and it gradually increased over the next 2 years.

Some have argued that the 2018 flare was a tidal disruption event (TDE) in an existing AGN \citep{trak19,Ricci_2020,Ricci2021,masterson2022}, while others propose a magnetic flux inversion scenario \citep{scepi21, laha2022}. Our recent study \citep{ghosh2023} covering the post-changing-look phase detected an emerging bright soft state with the soft X-ray flux reaching $\sim 5$ times that of the pre-flare value, while the hard X-rays and UV showed no significant new variation. 
1ES 1927+654 has an Eddington ratio\footnote{The Eddington ratio $\lambdaedd=\lbol/\ledd$ is the ratio between the measured bolometric luminosity $\lbol$ and the Eddington luminosity $\ledd$, where the latter is derived directly from the black hole mass.} (as measured on May 2023 using broadband spectral modeling of the UV-X-ray SED) of $\lambdaedd=0.23$  \citep{ghosh2023}.
This value of the Eddington ratio would likely place 1ES 1927+654 in the radiatively efficient, thin accretion disk regime (see, e.g., \citealt{Shakura_Sunyaev_1973, Novikov_Thorne_1973}), with typical radiative efficiencies of $\sim 10\%$. However, during the CL event (2018-2021) the source was mostly accreting at the Eddington rate \citep{Li2024}.

Our companion paper (Masterson et al. 2025, Nature, in press) finds significant short-term variability in the X-ray light curve and has detected a consistent quasi periodic oscillation (QPO) predominantly in the $2-10\kev$ band. The QPO was first detected in July 2022 with \xmm{} observations, exhibiting an 18-minute period. The QPO frequency corresponds to coherent motion on scales $<10 \, r_g$, where $r_g$ is the gravitational radius. The period decreased to 7.1 minutes over two years, from July 2022 to March 2024, with a decelerating period evolution. The QPO frequencies measured at the four epochs of \xmm{} observations are: $\nu = 0.93\pm 0.06$ mHz (July 2022), $\nu=1.67\pm 0.04$ mHz (Feb 2023), $\nu=2.21\pm 0.05$ mHz (Aug 2023), $\nu=2.35\pm 0.05$ mHz (Mar 2024). The QPO frequency was found to be strongly correlated with the X-ray spectral slope $\Gamma$ and the soft and hard X-ray fluxes. In addition, it was observed that the QPO is more prominently detected in the hard X-rays. The fractional root-mean-square (RMS) of the QPO increases from $2-7\%$ in the $0.3-2\kev$, to $15-20\%$ in the $2-10\kev$ band, across all observations. The origin of the QPO is still debated but several models such as stable mass transfer from a white dwarf companion, disk instabilities with a dependence on the accretion rate,
or magneto-acoustic oscillations in a variable corona have been extensively explored in Masterson et al. 2025.

Our second companion paper \citep{meyer2024} reports a new late-time radio flare in 1ES~1927+654, with a $\sim 40$- and 60-fold increase in the core radio flux density at 5 and 8.4 GHz, respectively, over a short period of about 6 months starting in early 2023. The study also detected spatially-resolved bipolar extensions of jets/outflows at $0.1-0.3\pc$  scales using Very Long Baseline Array (VLBA) observations. These outflows have been found to be mildly relativistic with a speed of $\sim 0.2c$. In Sections 3.1 and 3.2 of \cite{meyer2024}, we discuss the VLBI-resolution radio spectral energy distribution (SED) spanning from January 2020 to 2024 (Figure 3), along with core-subtracted K-band (23 GHz) VLBA images (Figure 4). These extended, resolved features provide clear evidence of a relativistic jet-driven outflow that is undergoing significant evolution, likely due to particle acceleration and shock interactions within the jet. The observed brightness temperature ($T_b$)\footnote{\begin{equation} \label{eq: Tb} T_b = \frac{4 \ln(2) c^2 S_\nu}{2 \pi k_B \theta_A \theta_B}\,\, {\rm K}, \end{equation} where $S_\nu$ is the specific intensity over the solid angle, and $\theta_A$ and $\theta_B$ are the full widths at half power of the major and minor axes of an elliptical Gaussian beam.} for the radio intensity values at 5 GHz, ranging from $10^{7}$~K to $10^{9}$~K, indicates extremely high-energy synchrotron-dominated emission. Such high brightness temperatures are typically associated with regions of intense, relativistic particle acceleration, suggesting that the emission is likely driven by a relativistic jet. Furthermore, it was observed that during the spectacular rise of the radio flare, the higher-frequency X band ($\sim 8$ GHz) became optically thin before the lower-frequency band ($\sim 5$ GHz), which is consistent with an optical depth from free-free absorption $\tau\propto \nu^{-2}$ .



We have continued to follow up 1ES 1927+654 with 
observations from space-based and ground-based missions to track the multi-wavelength evolution, and the recent soft X-ray rise, which we refer to as the `bright soft state' \citep{ghosh2023}. In this paper we report the most recent multiwavelength observations of the source from \swift{}, \xmm{}, \nustar{}, and {\it Zwicky Transient Facility} (ZTF) from May 2022 to April 2024. For clarity, in this paper we use `2018 flare' or `optical/UV flare' to indicate the original optical/UV outburst peaking in 2018, where `pre-flare' (alternatively pre-changing-look) phase of the source indicates anytime before 2017 (See Figure \ref{fig:timeline}). The `post-flare' period indicates the period after January 2022 (when the high X-ray state first subsided), and the `radio flare' refers to the recent major increase in GHz radio flux which began in early 2023 \citep{meyer2024}.

The paper is arranged as follows: Section \ref{sec:observations} discusses the observation, data reduction, and analysis. Section \ref{sec:results} lists the most important results. Section \ref{sec:discussion} discusses the main scientific topics, and Section \ref{sec:conclusions} lists the main conclusions.


\begin{table*}

\centering
  \caption{The details of multi-wavelength observations of 1ES~1927+654 used in this work. Refer to \citep{laha2022,ghosh2023} for all the previous \swift{} observations.  }\label{Table:obs1}
  \begin{tabular}{cccccccc} \hline\hline 

Observation band	& Telescopes			&observation date	&observation ID	& Exposure & Short-id	\\
			        &   				    & YYYY-MM-DD               &		        &(Sec)	\\ \hline 


X-ray and UV		& \xmm{}  & 2011-05-20 & 671860201 & 28649 & -- \\
"		            & \xmm{}  & 2022-07-26 & 902590201 & 27900 & -- \\
"		            & \xmm{}  & 2022-07-28 & 902590301 & 19000 & -- \\
"                   & \xmm{}  & 2022-07-30 & 902590401 & 19000 & -- \\
"                   & \xmm{}  & 2022-08-01 & 902590501 & 22000 & -- \\
"                   & \xmm{}  & 2023-02-21 & 915390701 & 34600 & -- \\
"                   & \xmm{}  & 2023-08-07 & 931791401 & 36400 & -- \\
"                   & \xmm{}  & 2024-03-04 & 932392001 & 33000 & -- \\
"                   & \xmm{}  & 2024-03-12 & 932392101 & 33100 & -- \\
\hline

X-ray       		& \nustar{} &2023-05-27	& 80902632002	&37117   &--	\\
"       		      & \nustar{} &2023-06-26	& 80902632004	&41074   &--	\\
"      		       & \nustar{} &2023-08-05	& 80902632006	&39840   &--	\\
"      		          & \nustar{} &2023-09-03	& 80902632008	&30987   &--	\\\hline

X-ray and UV  &{\it Swift-XRT/UVOT}&2018-05-17	& 00010682001	&2190   &S01	\\\hline


''		            & ''                 &2023-05-09		& 00010682079	&710   &S79	\\
''		            & ''                 &2023-05-13		& 00010682080	&885   &S80	\\
''		            & ''                 &2023-05-17		& 00010682081	&754    &S81	\\
''		            & ''                 &2023-05-21		& 00010682082	&887    &S82	\\
''		            & ''                 &2023-05-20		& 00010682083	&932   &S83	\\
''		            & ''                 &2023-05-22		& 00010682084	&993   &S84	\\
''		            & ''                 &2023-05-23		& 00010682086	&942    &S86	\\
''		            & ''                 &2023-05-25		& 00010682087	&436   &S87	\\
''		            & ''                 &2023-05-26		& 00010682088	&433   &S88	\\
''		            & ''                 &2023-06-09		& 00010682089	&1572   &S89	\\
''		            & ''                 &2023-06-11		& 00010682090	&953    &S90	\\
''		            & ''                 &2023-06-12		& 00010682091	&867    &S91	\\
''		            & ''                 &2023-06-13		& 00010682092	&872   &S92	\\
''		            & ''                 &2023-06-15		& 00010682094	&1624    &S94	\\
''		            & ''                 &2023-06-17		& 00010682095	&1654   &S95	\\
''		            & ''                 &2023-06-21		& 00010682096	&1424    &S96	\\
''		            & ''                 &2023-06-24		& 00010682097	&985    &S97 	\\
''		            & ''                 &2023-06-27		& 00010682098	&827    &S98 	\\
''		            & ''                 &2023-07-03		& 00010682100	&1008    &S100 	\\
''		            & ''                 &2023-07-06		& 00010682101	&930   &S101 	\\
''		            & ''                 &2023-07-09		& 00010682102	&907    &S102 	\\
''		            & ''                 &2023-07-15		& 00010682103	&852    &S103 	\\
''		            & ''                 &2023-07-25		& 00010682104   &366    &S104	\\
''		            & ''                 &2023-07-27    	& 00010682105	&393    &S105 	\\
           \hline
\end{tabular}  

\end{table*}

\begin{table*}

\centering
  \caption{The details of multi-wavelength observations of 1ES~1927+654 used in this work. }\label{Table:obs2}
  \begin{tabular}{cccccccc} \hline\hline 
Observation band	& Telescopes			&observation date	&observation ID	& Net exposure & Short-id	\\
			        &   				    & YYYY-MM-DD               &		        &(Sec)	\\ \hline 

X-ray and UV            & {\it Swift-XRT/UVOT}

  &2023-08-02		& 00010682106   &822    &S106 	\\
''		            & ''                 &2023-08-05		& 00010682107	&812    &S107 	\\
''		            & ''                 &2023-08-08		& 00010682108	&750    &S108	\\
''		            & ''                 &2023-08-11		& 00010682109	&1012    &S109 	\\ 

'' & '' &2023-08-14		& 00010682110	&963    &S110 	\\
''		            & ''                 &2023-08-17		& 00010682111	&872    &S111 	\\
''		            & ''                 &2023-08-26		& 00010682113	&908    &S113 	\\
''		            & ''                 &2023-09-23		& 00010682115	&850    &S115	\\
''		            & ''                 &2023-09-27		& 00010682116	&938    &S116 	\\
''		            & ''                 &2023-10-01		& 00010682117	&1171    &S117	\\
''		            & ''                 &2023-10-05		& 00010682118	&807    &S118	\\
''		            & ''                 &2023-10-09		& 00010682119	&880    &S119	\\
''		            & ''                 &2023-10-13		& 00010682120	&948    &S120 	\\
''		            & ''                 &2023-10-17		& 00010682121	&920    &S121  \\
''		            & ''                 &2023-10-21		& 00010682122   &862    &S122 	\\
''		            & ''                 &2023-10-24		& 00010682123	&867    &S123 	\\
''		            & ''                 &2023-11-06		& 00010682124   &687    &S124 	\\
''		            & ''                 &2023-11-10		& 00010682125	&842    &S125 	\\
''		            & ''                 &2023-11-15		& 00010682126	&717    &S126 	\\
''		            & ''                 &2023-11-18		& 00010682127	&868    &S127 	\\
''		            & ''                 &2023-11-22		& 00010682128	&935   &S128	\\
''		            & ''                 &2023-11-26		& 00010682129	&1096   &S129	\\
''		            & ''                 &2023-11-30		& 00010682130	&1045    &S130	\\
''		            & ''                 &2023-12-04		& 00010682131	&1278   &S131	\\
''		            & ''                 &2023-12-08		& 00010682132	&892   &S132	\\
''		            & ''                 &2023-12-13		& 00010682133	&557    &S133	\\
''		            & ''                 &2023-12-25		& 00010682134	&883   &S134	\\
''		            & ''                 &2023-12-28		& 00010682135	&933   &S135	\\
''		            & ''                 &2024-01-05		& 00010682137	&933   &S137	\\
''		            & ''                 &2024-01-09		& 00010682138	&852    &S138	\\
''		            & ''                 &2024-01-13		& 00010682139	&908    &S139	\\
''		            & ''                 &2024-01-17		& 00010682140	&996   &S140	\\
''		            & ''                 &2024-01-24		& 00097196023	&797   &S140A	\\
''		            & ''                 &2024-02-03		& 00097196024	&847   &S140B	\\
''		            & ''                 &2024-02-21		& 00097196025	&872   &S140C	\\
''		            & ''                 &2024-02-28		& 00010682141	&960   &S141	\\
''		            & ''                 &2024-03-10		& 00097196027	&1020   &S141A	\\
''		            & ''                 &2024-04-06		& 00016519001	&847   &S143	\\
''		            & ''                 &2024-04-11		& 00016519002	&900   &S144A	\\

\hline 
\end{tabular}  
\end{table*}

\section{Observations and data analysis} \label{sec:observations}
In this section, we discuss the X-ray, UV and optical observations used in this work, which involves archival observations, as well as others obtained through the Director's Discretionary Time (DDT) and Guest Investigator (GI) programs. Tables \ref{Table:obs1} and \ref{Table:obs2} list the observation details. For details of the radio data presented in this paper, see \cite{meyer2024}.

\subsection{XMM-Newton}\label{subsec:XMM}

We have analyzed a total of nine \xmm{} \citep{2001A&A...365L...1J} observations taken in 2011, 2022, 2023 and 2024 encompassing the pre-changing look (pre-CL) and the recent soft X-ray rise. The first May-2011 observation was taken during the pre-CL/pre-flare state of the source (See Table~\ref{Table:obs1} and Figure \ref{fig:timeline} for details), and the one in Feb-2023 (DDT; PI-S.Laha) was taken exactly when the source started to show signs of a radio burst and the other Aug-2023 (DDT; PI-S.Laha) observation taken when the source has reached the radio-loud phase. The observations in March 2024 were obtained through DDT (PI: M. Masterson).

We used the latest XMM–Newton Science Analysis System (SAS v19.0.0) to process the Observation Data Files (ODFs) from all observations.  For brevity we report only the European Photon Imaging Camera \citep[EPIC-pn][]{2001A&A...365L..18S} observations in this paper. The {\tt EVSELECT} task was used to select the single and double events for the pn detector ({\tt PATTERN<4}). We created light curves from the event files for each observation to account for the high background flaring using a rate cutoff of $<0.4 \rm \, count\, s^{-1}$. We found significant particle background flares in the Aug-2023 observations, but lesser in other epochs. We found that the Feb 2023, Aug 2023 and March 2024 observations show some pile up in the soft X-rays (using the SAS task {\tt epatplot}). For these observations we selected an annular region for source photons with an inner and outer radii of $8$ arcsec and $30$ arcsec respectively, centered on the source. For the other observations the source regions were extracted from circular regions of 40 arcsec centred on the source. In all cases the background photons were extracted from appropriate regions away from the source but on the same CCD. The response matrices were generated using the SAS tasks {\tt arfgen} and {\tt rmfgen}. The spectra were grouped using the command {\tt ftgrppha} with a minimum of 20 counts in each energy bin. 

  We analyzed the EPIC-pn spectra for the nine epochs separately using the phenomenological models (as per XSPEC notation) {\tt tbabs*ztbabs*(powerlaw+bbody)}. The black-body component {\tt bbody} was required to model the soft X-ray excess while the {\tt powerlaw} model is used to describe the Comptonized non-thermal spectrum from AGN corona. The {\tt tbabs} model was used to describe the Galactic absorption, with a column density, $N_H$ held fixed to $6.42 \times 10^{20}$~cm$^{-2}$ \citep{kalberla2005}. The {\tt ztbabs} model was used to describe the  host-galaxy intrinsic absorption, with a best fit column density $\nh \sim 4-5 \times 10^{20}$~cm$^{-2}$. This low level of intrinsic absorption has been present since the pre-CL state (in 2011) indicating a large-scale absorber, possibly galactic dust lanes \citep[see for e.g.,][]{laha2020}. We have also used two Gaussian profiles to model the narrow emission lines at $\sim 0.56\kev$ and $\sim 1\kev$ in the soft X-ray $0.3-2\kev$ spectra. To accurately constrain and characterise the narrow emission lines in the soft X-rays we require a simultanoeus spectral fit of EPIC-pn and the high resolution Reflection Grating Spectrometer (RGS). This is beyond the scope of this paper, and therefore we defer the study of the emission lines to a future work. The best fit continuum parameters are reported in Table \ref{Table:xmm_pnfit}.


\subsection{Swift}
\subsubsection{Swift-XRT and UVOT}\label{subsubsec:swift_reprocess}

1ES~1927+654 was observed by The Neil Gehrels \emph{Swift} Observatory X-ray Telescope \citep[XRT,][]{burrows2005} initially on a monthly cadence from January 2022 to November 2022, and then at a weekly and bi-weekly cadence from December 2022 to May 2024 (See Tables~\ref{Table:obs1} and \ref{Table:obs2}) under DDT and \emph{Swift}-GI programs (PI: S.Laha). We use abbreviations for the \swift{} observations to easily identify them. The names begin with the letter ``S" and have been used in our earlier works \citep{laha2022, ghosh2023}. The observations up to S78 (May 2018- May 2022) have been reported in our previous works \cite{laha2022, ghosh2023}. In this paper we discuss only the observations from May 2022 to April 2024 (S79-S144A).

We followed the automated XRT analysis approach via the online tools\footnote{https://www.swift.ac.uk/user-objects} \citep{Evans2009_xrt} for the XRT data in all our observations as recommended for point sources by the \swift{} help desk. We used similar spectral models to fit the XRT spectra as have been used in \xmm{} observations (See Tables \ref{Table:swift_obs1} and \ref{Table:swift_obs2}). We note here that due to the poor statistical quality of the XRT spectra, mostly due to the low effective area, and low exposure times per snapshot (300-800 sec), resulting in low count rate $\sim 0.8$ counts/sec (compared to $\sim 15$ counts/sec in \xmm{}) we did not statistically require the intrinsic absorber model component ({\tt ztbabs}), which was needed in the \xmm{} spectral fit. In other words, we did not find any improvement in fit statistics ($\Delta\chi^2=0$) on addition of this model in the XRT spectra. Not being able to model the intrinsic absorption with the XRT spectra results in a slightly lower estimation of the intrinsic soft X-ray $0.3-2\kev$ flux by $15-20\%$ compared to \xmm{}, but the $2-10\kev$ flux remains unaffected.

\swift{} UVOT~\citep{2005SSRv..120...95R} observed the source simultaneously along with \swift{}-XRT. Refer to \cite{ghosh2023} for a full description of UVOT data reprocessing and analysis, which we follow here. The UV flux densities were corrected for Galactic absorption using the correction magnitude of $\rm A_{\lambda}=0.690$ obtained from the NASA Extragalactic Database (NED\footnote{https://ned.ipac.caltech.edu}).
Figure \ref{fig:xray_uv_alpha_ox_zoomed} shows the X-ray flux and UV flux-density light curves as measured by \swift{}.

 We estimated the Eddington ratio of the source, as on April 2024, using the relation $\lambdaedd=\lbol/\ledd$, where $\ledd$ is the Eddington luminosity of the
source, assuming a BH of mass $\sim 10^6\msol$ \citep{Li_broad_line_1es}. To estimate $\lbol$ we have used the integrated UV luminosity in the band $0.001-100\ev$, and the X-ray luminosity in the $0.3-10\kev$, obtained by fitting the Swift-UVOT photometric data simultaneously with the XRT spectrum (see \citealt{ghosh2023} for details). We estimate an Eddington ratio of $\lambdaedd \sim 0.3$. Considering a $2-10\kev$ luminosity of $4\times 10^{42}\lunit$, this would give us a bolometric correction of $\kappa\sim \lbol/{L_{2-10\kev}\sim 10}$ consistent with typical sub-Eddington AGN \citep{vasudevan2007}.

\subsection{NuSTAR}

1ES 1927+654 has been observed by \texttt{NuSTAR} four times in 2023 (GI program: PI- S.Laha) when the source was getting radio bright, capturing the hard X-ray spectra during the radio rise. The data were processed using the \texttt{NuSTAR} Data Analysis Software (NUSTARDAS) version 2.1.2. Calibration of the raw event files were performed using the \textit{nupipeline} script and the response file from \texttt{NuSTAR} Calibration Database (CALDB) version 20211020. 
The source and background spectra are extracted from $30''$ ($\approx 50\%$ of the encircled energy fraction--EEF at 10 keV) and $50''$ circular regions, respectively. The \textit{nuproducts} scripts are used to generate the source and background spectra files, along with response matrix files (RMF) and ancillary response files (ARF). Finally, using \textit{grppha}, the \texttt{NuSTAR} spectra are grouped with at least 20 counts per bin in order to use the $\chi^2$ statistics. 

Figure \ref{fig:overplot_nustar_xmm} shows the four epochs of \nustar{} spectra along with the \xmm{} spectra in 2023. The \nustar{} $3-40\kev$ spectra are well described by a simple power law with a slope 
$\Gamma\sim 3.2$ consistent with the four observations (and also consistent with \xmm{} observations during similar epochs). See Table \ref{Table:nustar_table} for the best fit parameters and flux values.

\subsection{Zwicky Transient Facility}

\begin{figure*}[h!]
    \centering
    \includegraphics[height=19cm,width=16.5cm]{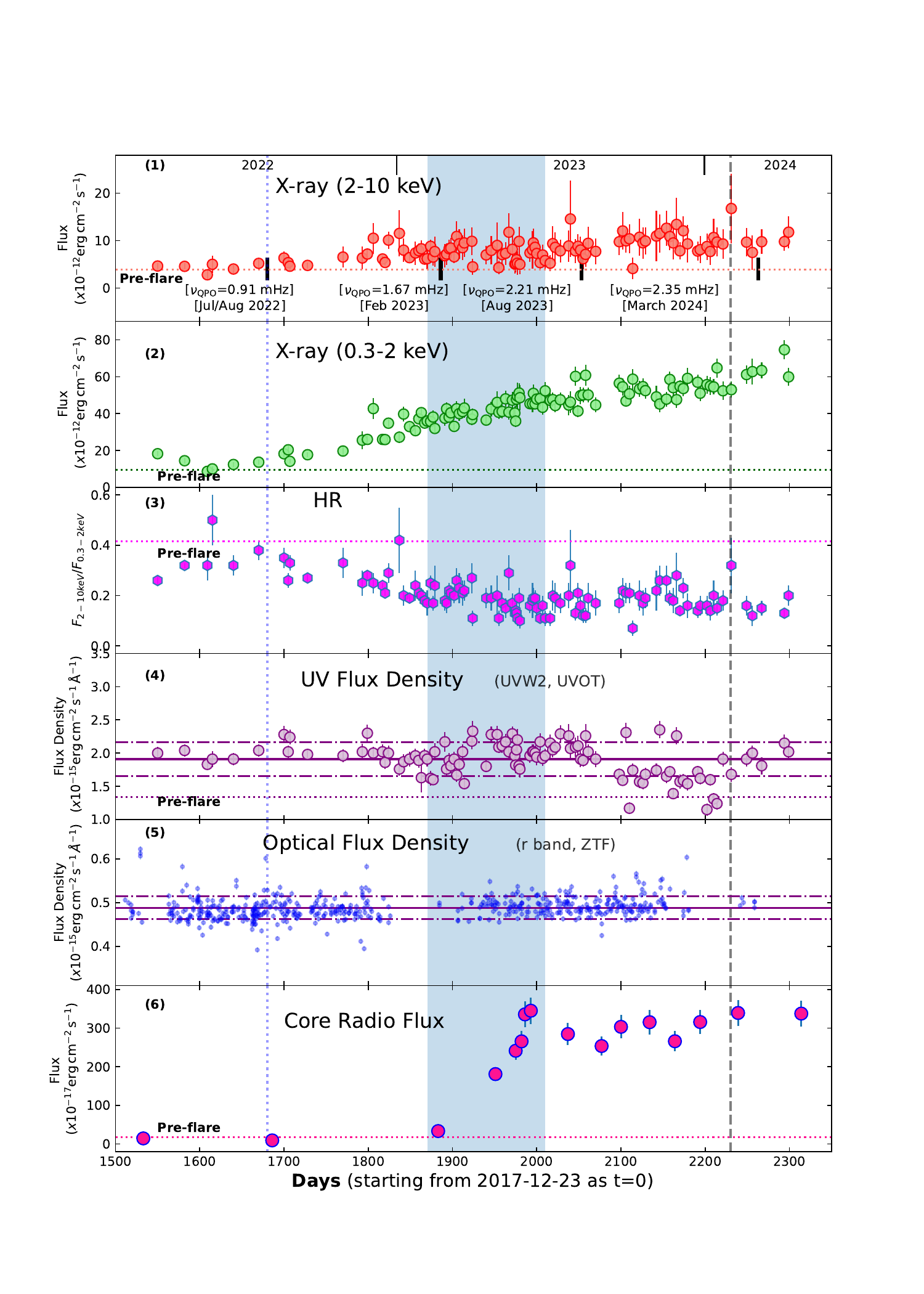}
    \caption{\textbf{The X-ray, UV, optical, and radio light curves of 1ES1927+654 during the radio rise, soft X-ray rise, and the QPO detection phase (May 2022 - April 2024).} See Table \ref{Table:swift_obs1}, \ref{Table:swift_obs2}, for details. The start date of the light curve is 2017-12-23, corresponding to the burst date reported by \cite{trak19}. The blue shaded region correspond to the time when we detected the exponential rise in the radio (5GHz) flux. The radio flux increased by a factor of almost $\sim 60$ in a matter of a few months. The blue dotted line on the left corresponds to the time (May 2022) when the soft X-ray started to rise, the QPO detected and the radio jet was formed. The dashed black line on the right corresponds to the time (Feb 2024) when we detected spatially resolved jets at $0.1-0.3 \pc$ scales. The dotted horizontal lines in every panel refer to their pre-flare values (as in May 2011). From top to bottom panels are: (1) The $2-10\kev$ X-ray flux (\swift{}-{XRT}), (2) The $0.3-2\kev$ X-ray flux (\swift{}-{XRT}), (3) The hardness ratio: $F_{2-10\kev}/F_{0.3-2\kev}$, (4) The UV (UVW2) flux density in units of $10^{-15}\funita$ (\swift{}-{UVOT}), (5) The Optical r-band (ZTF) flux density (in units of $10^{-15}\funita$),  and (6) the Core radio flux at $<1\pc$ spatial resolution ({\it VLBA}, 5GHz) as reported in \cite{meyer2024}. The top panel (panel 1) has four vertical lines corresponding to the time when the QPO was detected in the $2-10\kev$ band using \xmm{} observations. The QPO frequencies and the date of observations are listed in panel 1. The X-ray flux in units of $ 10^{-12}\funit{}$ is corrected for Galactic absorption. } 
    \label{fig:xray_uv_alpha_ox_zoomed}
\end{figure*}

We obtained the Zwicky Transient Facility (ZTF) optical photometric data for 1ES~1927+654 for the period starting Jan 2018 to March 2024 through the public release website DR 21\footnote{\url{https://www.ztf.caltech.edu/ztf-public-releases.html}}  \citep{graham_ztf_2019,bellm_ztf_2019}. The ZTF survey covers the visible northern sky, reaching median depths of approximately 20.8 magnitudes in the g band and 20.6 magnitudes in the r band (AB, 5$\sigma$) with 30-second exposures. It utilizes point spread function fitting for photometry to construct light curves. We plot the ZTF-optical r-band light curve of 1ES~1927+654, roughly at a daily cadence containing observations until March 01, 2024, in Figure \ref{fig:xray_uv_alpha_ox_zoomed}. Note that the host galaxy contribution is not subtracted from the r-band flux.

\subsection{Fermi/LAT}
The \textit{Fermi}/LAT gamma-ray observatory, launched in 2008, is sensitive to photons in the energy range of $\sim$30 MeV to 300 GeV and mostly operates in a sky-scanning mode which has produced a deep archive of gamma-monitoring data covering the last 16 years \citep{fermi}.

\textit{Fermi}/LAT event and spacecraft data were extracted using a 15$^\circ$ region of interest (ROI), an energy range of 100 MeV - 300 GeV, a zenith angle cut of 90$^\circ$, and the recommended event class and type for point source analysis. The time cuts of the collected data included the entire \textit{Fermi} mission runtime which corresponds to a mission elapsed time range (MET) of 239557417 - 741587557. The analysis was performed using the publicly available \texttt{easyFermi} which processes the user provided data files to perform a \textit{Fermi}/LAT binned likelihood analysis following the standard methodology. The \texttt{easyFermi} analysis was performed over a time range from January 1, 2024 at 12AM to June 1, 2024 at 12AM corresponding to a MET time range of 725760001 - 738892801. The galactic model \texttt{gll\_iem\_v07.fits} and the isotropic diffuse emission model \texttt{iso\_P8R3\_SOURCE\_V3\_v1.txt} were used for the analysis. To obtain a converged fit, a free source radius of 10$^\circ$ was used with only the normalization free. 

We produced a lightcurve with twenty time bins over the above MET range.  Based on the values of the test statistic (TS)\footnote{test statistic (TS) is a maximum-likelihood method commonly used as a {\it Fermi} data analysis tool to determine if a source is detected or not.} for the lightcurve periods, at no time period is the source significantly detected, with the highest TS being 0.665 associated with the MET time range 733639750 - 734296390. No detectable flare was measured in the \textit{Fermi} data during the analysis time period. 

In addition to the above analysis, we also produced a lightcurve for the position of 1ES~1927+654 in a similar way for the entire operating time of Fermi through early 2024 with a 6-month time binning. Again no significant detection was made at any time, with the highest value of the TS $\sim$4 (likely a background or nearby source fluctuation) and the vast majority consistent with zero. 


\begin{figure*}[]
 \hbox{
\includegraphics[height=6cm, width=9cm] {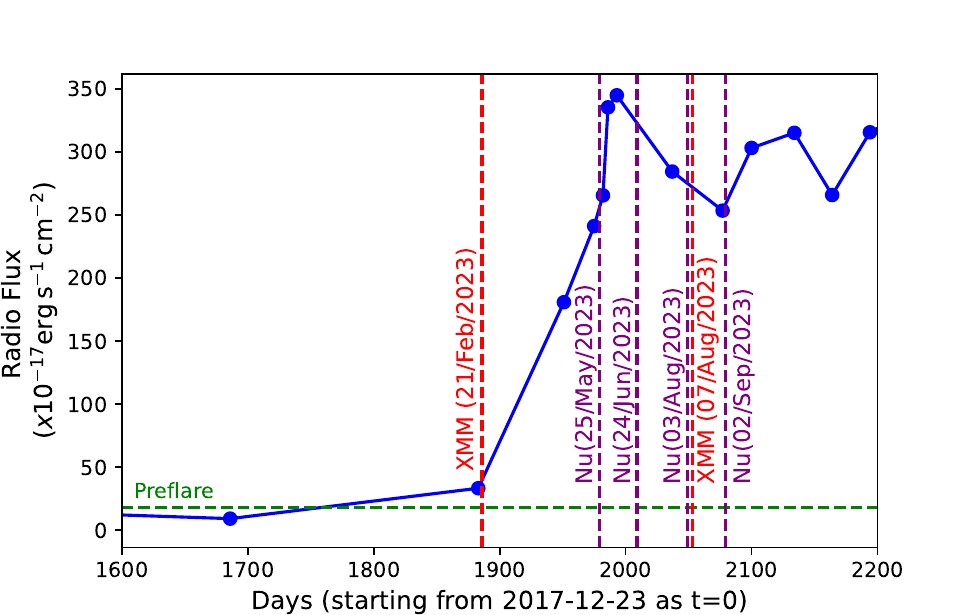}
\includegraphics[height=6cm, width=9cm]{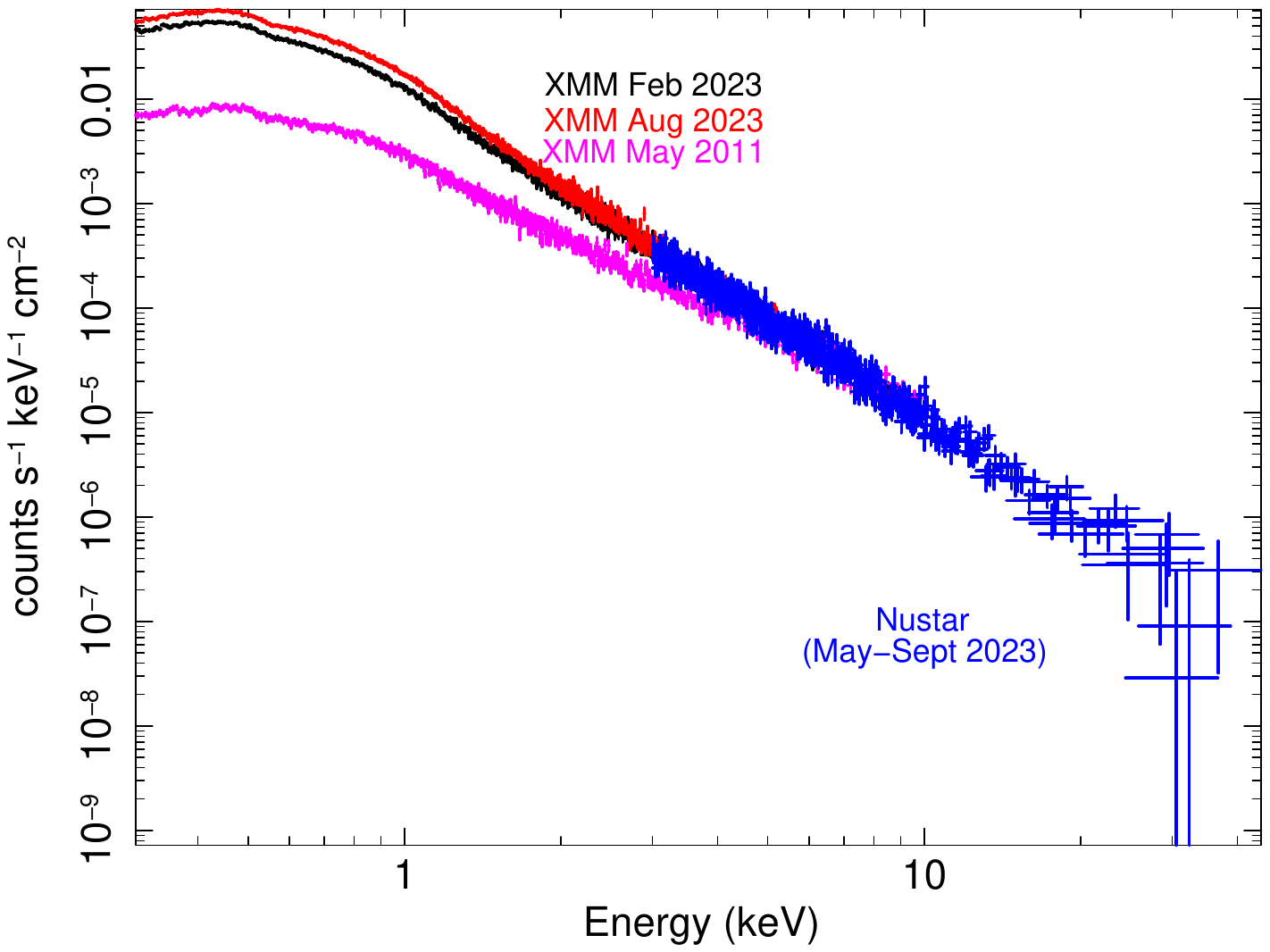}
}
\caption{\textbf{The \xmm{} and \nustar{} observations of 1ES~1927 coincident with the radio-flaring phase (Feb-Sept 2023).} {\it Left panel:} The radio light curve during the outburst, with the vertical lines denoting the X-ray observations with \xmm{} (XMM) and \nustar{} (Nu). We note that the radio-rise and the plateau phase was well covered by X-ray observations. {\it Right:} The three \xmm{} spectra in the $0.3-10\kev$ band from the three epochs (pre-flare:May 2011 in pink, rise-of-radio:Feb 2023 in black, high-radio:Aug 2023 in red), along with the \nustar{} spectra in blue from the four epochs during the radio rise (May, June, Aug, Sep 2023). We clearly note that the $2-10\kev$ spectral flux and slope $\Gamma$ have been non-variable ($<5\%$) during the radio-rise phase. See Tables \ref{Table:xmm_pnfit} and \ref{Table:nustar_table} for details. }
\label{fig:overplot_nustar_xmm}
\end{figure*}


\section{Results}
\label{sec:results}
We summarize the multi-wavelength results from our campaign during the period May 2022 - April 2024. The \xmm{} and \nustar{} X-ray observations were coordinated to cover the time period of the radio flare (Feb-Aug 2023). Fig \ref{fig:overplot_nustar_xmm} left panel shows the time stamps when the X-ray observations were made covering the radio flare phase. Fig \ref{fig:overplot_nustar_xmm} right panel shows the \xmm{} and \nustar{} spectral over-plots covering this time period. The \nustar{} spectra in blue do not show any variation in slope or flux at $<5\%$ during the $\sim 60$ times radio flux rise. See Tables \ref{Table:xmm_pnfit} and \ref{Table:nustar_table} for details.

In this section we clearly make a distinction between the two quantities: The soft X-ray excess (soft-excess from now on) and the soft X-ray fluxes. The former is the `excess' in the $0.3-2\kev$ flux after fitting a power law continuum and we have used a black body model to describe it. The latter on the other hand refers to the total flux in the $0.3-2\kev$ band including contributions from both the power law and the black body. We note that the soft-excess contributes to around $\sim 30-40\%$ of the soft X-ray flux.


\subsection{The X-ray light curve evolution}

{Both the soft X-ray (0.3-2 keV) and hard X-ray ($2-10\kev$) flux of 1ES~1927+654 have shown notable variability on week-to-month timescales in the May-2022 to Apr-2024 period. As previously reported in \cite{ghosh2023}, the \swift{} XRT observations of 20th May 2022 showed that the $0.3-2\kev$ and $2-10\kev$ fluxes had reached their pre-flare values of $\sim 4\times 10^{-12}\funit$ and $\sim 9\times 10^{-12}\funit$ respectively. As shown in Figure~\ref{fig:xray_uv_alpha_ox_zoomed} and Tables \ref{Table:swift_obs1} and \ref{Table:swift_obs2}, the soft X-ray flux has since risen steadily over the past two years.  As of 11th April 2024, the soft X-ray flux level is $\sim 8$ times that of its pre-flare value. On the other hand, the $2-10\kev$ light curve exhibits stochastic variations ($<2$ times) with a slow steady rise. These estimates are also corroborated by \xmm{} observations during this time period (See Table \ref{Table:xmm_pnfit}). We also note that the soft-excess flux increased by a factor of $\sim 10$ during this period (See Table \ref{Table:xmm_pnfit}).

\subsection{X-ray spectroscopy}

As discussed in Sec \ref{subsec:XMM}, we analyzed nine epochs of \xmm{} EPIC-pn spectra with phenomenological models and the detailed results are quoted in Table \ref{Table:xmm_pnfit}. We find that the spectral index steepens from May 2011, i.e pre-changing-look epoch ($\Gamma \simeq 2.44$), to the pre-radio and soft X-ray flare epoch ($\Gamma \simeq 2.6$--2.8) in May 2022, reaching $\Gamma \simeq 3.2$--3.3 during the radio flare. Such steep spectral indices are inconsistent with what is typically observed in Seyfert galaxies \citep{ricci2017} or quasars \citep{zappacosta2023}. Most of the increase in the soft-excess flux is attributed in this (black-body) model to the growth of the normalization of the thermal component and not in its temperature. Remarkably, the temperature of the blackbody component remains stable within a rather small range, $kT_{\rm bb} \sim 0.13-0.16\kev$. A similar spectral model was applied to the short ($\sim 300-1500$ sec) \swift{} spectra, yielding very similar results. See Tables  \ref{Table:swift_obs1} and \ref{Table:swift_obs2}.

We do not detect any statistically significant ionized absorption in any of the EPIC-pn spectra. We do not find any trace of a FeK$\alpha$ emission line at $\sim 6.4\kev$ at any of the epochs, consistent with previous studies \citep{Ricci_2020,Ricci2021,masterson2022}.  The 90\% upper limit on the equivalent width of a K$_{\alpha}$ fluorescent line from neutral Fe is $\simeq 120\ev$.

A steep spectrum is confirmed by the four NuSTAR observations taken during the radio flare in May-Sept 2023 (Tab.~\ref{Table:nustar_table}). We employed a model constituted by a photoelectrically absorbed power-law. The model fits the data well. It confirms a steep spectral index ($\Gamma$~=~3.1--3.2) over the wider energy range of the NuSTAR instruments ({\it i.e.}, up to $\simeq$30~keV). If a self-consistent Comptonization model is used ({\tt nthcomp} in XSPEC) the electron temperature is well constrained to $kT_{\rm e}=8.4^{+8.8}_{-2.6}\kev$, in a joint simultaneous fit of all the four \nustar{} observations.

\subsection{The optical and UV light curve}
The UV flux density of 1ES~1927+654 was monitored using the \swift{} UVOT, and we quote the UVW2 band ($1928 \, \rm \angstrom$) for consistency with our earlier works. The UV in this time period shows very weak variability, not more than $30\%$ of the mean value (See Table \ref{Table:swift_obs1} and Fig \ref{fig:xray_uv_alpha_ox_zoomed}.  Thus the UV variability is not significant. This may indicate a near-steady rate of accretion. We also plot the optical r-band light curve from ZTF in Fig. \ref{fig:xray_uv_alpha_ox_zoomed} and find only minimal fluctuations at days timescale. We confirm that there is no optical or UV flaring during this time period.

\begin{table*}
\centering
 \caption{The best-fit parameters obtained from fitting the \xmm{} EPIC-pn observations. \label{Table:xmm_pnfit}}
{\renewcommand{\arraystretch}{1.2}
\setlength{\tabcolsep}{1.7pt}
\begin{tabular}{cccccccccccccccccccccccc} \hline\hline
Models 		& Parameter 	&20/05/11 		&26/07/22 		&28/07/22 		&30/07/22 		&01/08/22 		&21/02/23		&07/08/23 		&04/03/24         &12/03/24		   \\ \hline 

Gal. abs.  	& $\nh \,(\times 10^{20})$ & $ 6.4$ (f) &$ 6.4$ (f)  & $ 6.4$ (f)  & $ 6.4$ (f)  & $ 6.4$ (f)  & $ 6.4$ (f)  & $ 6.4$ (f)  & $ 6.4$ (f)  & $ 6.4$ (f)  \\

ztbabs  	& $\nh \,(\times 10^{20})$ & $4.04^{+0.53}_{-0.52}$ & $3.22^{+0.70}_{-0.67}$ & $2.39^{+0.65}_{-0.64}$ & $2.11^{+0.68}_{-0.69}$ & $1.02^{+0.86}_{-1.08}$ & $4.69^{+0.56}_{-0.56}$ & $5.60^{+0.52}_{-0.52}$ & $5.91^{+0.64}_{-0.64}$ & $5.76^{+0.55}_{-0.54}$  \\

black-body  & $T_{in}$ ($\kev$)  &$0.16^{+0.01}_{-0.01}$ & $0.15^{+0.01}_{-0.01}$ & $0.15^{+0.01}_{-0.01}$ & $0.14^{+0.01}_{-0.01}$ & $0.15^{+0.01}_{-0.01}$ & $0.14^{+0.01}_{-0.02}$ & $0.14^{+0.01}_{-0.01}$ & $0.15^{+0.01}_{-0.01}$ & $0.15^{+0.02}_{-0.02}$ \\
            & norm  ($10^{-4}$)  &$0.39^{+0.04}_{-0.03}$ & $0.72^{+0.12}_{-0.09}$ & $0.66^{+0.10}_{-0.08}$ & $0.61^{+0.08}_{-0.07}$ & $0.79^{+0.10}_{-0.09}$ & $2.97^{+0.17}_{-0.16}$ & $4.88^{+0.18}_{-0.17}$ & $5.59^{+0.22}_{-0.21}$ & $5.39^{+0.18}_{-0.17}$   \\

 powerlaw 	& $\Gamma$ & $2.45^{+0.04}_{-0.04}$ & $2.65^{+0.04}_{-0.04}$ & $2.70^{+0.04}_{-0.04}$ & $2.71^{+0.04}_{-0.04}$ & $2.48^{+0.06}_{-0.07}$ & $3.10^{+0.04}_{-0.04}$ & $3.21^{+0.04}_{-0.04}$ & $3.26^{+0.05}_{-0.05}$ & $3.27^{+0.04}_{-0.04}$ \\
 
            & norm ($10^{-2}$)&$0.28^{+0.01}_{-0.01}$ & $0.45^{+0.02}_{-0.02}$ & $0.49^{+0.02}_{-0.02}$ & $0.51^{+0.02}_{-0.02}$ & $0.42^{+0.02}_{-0.03}$ & $1.07^{+0.05}_{-0.05}$ & $1.36^{+0.06}_{-0.05}$ & $1.49^{+0.07}_{-0.07}$ & $1.45^{+0.06}_{-0.06}$ \\ \hline
            
            &$\rm $$\rm \chi^2/dof$  &$688/695$	&$751/762$ &$643/624$ &$588/639$ &$929/950$	 &$607/632$    &$657/652$  &$623/626$ &$733/680$\\ \hline

            & $L^{bb}_{(0.3-2)\kev}$ ($10^{42}$)&$2.18^{+0.11}_{-0.11}$ & $2.41^{+0.22}_{-0.22}$ & $2.46^{+0.22}_{-0.22}$ & $2.59^{+0.20}_{-0.20}$ & $2.95^{+0.21}_{-0.20}$ & $13.06^{+0.43}_{-0.43}$ & $21.74^{+0.49}_{-0.49}$ & $25.94^{+0.62}_{-0.61}$ & $25.06^{+0.50}_{-0.50}$ \\ \hline

            & $L^{pow}_{(2-10)\kev}$ ($10^{42}$) &$2.45^{+0.03}_{-0.03}$ & $2.89^{+0.03}_{-0.03}$ & $3.01^{+0.03}_{-0.03}$ & $3.11^{+0.04}_{-0.04}$ & $3.33^{+0.04}_{-0.04}$ & $3.95^{+0.05}_{-0.05}$ & $4.35^{+0.05}_{-0.05}$ & $4.48^{+0.06}_{-0.06}$ & $4.29^{+0.05}_{-0.05}$ \\ \hline






\end{tabular} } 
 { The unabsorbed soft X-ray black-body ($0.3-2\kev$) and powerlaw ($2-10\kev$) luminosities denoted by $L^{bb}_{(0.3-2)\kev}$ and $L^{pow}_{(2-10)\kev}$ respectively are in the units of $\lunit$}\\

\end{table*}

\subsection{Estimating the total energy in the $0.3-2\kev$, $2-10\kev$ and the UV bump.}

We estimated the total energy pumped into the system during the time interval May 2022 to May 2024 by integrating the area under the light curve for the soft ($0.3-2\kev$) and hard bands ($2-10\kev$). This was done using the python scipy function {\tt integrate.simps}. The UV energy was estimated by assuming a diskbb of a temperature of kT$=5\ev$ \citep{ghosh2023}, and the normalization was matched at the wavelength 2200 Angstrom with the measured value of the UVW2 filter. We used the mean value of the UVW2 flux density (as in Fig \ref{fig:xray_uv_alpha_ox_zoomed}) and assumed to be constant while integrating (as it did not vary beyond $30\%$).  The total integrated energy under the \swift{}-XRT and UVOT light curves in the $0.3-2\kev$, the $2-10\kev$ and the UV bands are $1.56\times 10^{51}$ erg, $3.12\times 10^{50}$ erg, and $5.76\times 10^{50}$ erg, respectively. We clearly find that the total X-ray energy surpasses the energy in the UV, indicating that there may be some other source of energy than that of the UV accretion.

\subsection{The correlations}

 We used the python {\tt scipy} function {\tt spearmanr} to estimate the Spearman rank correlation between the hard/soft X-ray and UV fluxes. The Spearman rank coefficient measures the strength of the correlation, and the { p-value} measures how likely the correlation is not due to chance. For example, a small p-value (typically $<0.01$) indicates that we can reject the null hypothesis (i.e., no correlation) and conclude that there is a statistically significant relationship between the variables. Figures \ref{fig:swift_correlations} and \ref{fig:xmm_correlations} show the correlations and their corresponding coefficients and the p-values. In Figure \ref{fig:swift_correlations}, involving weekly/bi-weekly cadence flux values obtained using \swift{} observations from May 2022 - April 2024,  we do not find any significant correlation between the parameters (UV flux density, soft-excess and the $2-10\kev$ flux). A Spearman Rank correlation test ruled out any correlation at 99.99\% confidence. 

On the other hand, the observations from \xmm{} taken sparsely (once in 6 months) over the same time period is ideal to track the longer term evolution of these quantities. Figure \ref{fig:xmm_correlations} left panel shows that there is a strong correlation/trend between the power law slope and the $2-10\kev$ luminosity, and the right panel shows a gradual trend of an increase in the $2-10\kev$ luminosity with the black-body (soft-excess) luminosity. The red data points in each of the panels denote the values obtained from May-2011 \xmm{} observation when the source was in quiescence (pre-changing-look), and is not included in the correlation calculations. As a caveat we note that in Figure \ref{fig:xmm_correlations} right panel although there is a trend of both the soft-excess and power law luminosity increasing simultaneously, the amount of increase in the $2-10\kev$ luminosity is smaller ($<2$ times) compared to the increase in the soft-excess ($\sim 10$ times). See Table \ref{Table:xmm_pnfit} for details.


\begin{table*}
\fontsize{8}{10}
\centering
  \caption{The spectral parameters obtained using {\it Swift} and \xmm{} UV and X-ray observations of 1ES~1927+654. For comparison with the pre-flare values, we keep the \xmm{} observation. \label{Table:swift_obs1}} 
  \begin{tabular}{cccccccccccc} \hline\hline 

ID(DD/MM/YY)			&$F_{0.3-2\kev}^{\rm A}$	&$F_{2-10\kev}^{\rm A}$ &$F_{1.5-2.5\kev}^{\rm A}$  & kT (keV) & $\Gamma$ & UV filter& UV flux density$^{\rm B}$ &$\alpha_{\rm OX}$ & $\rm \chi^2/ \chi^2_{\nu}$ \\ \hline  

X1 (20/05/11)	    &   $9.41\pm0.66$   & $3.92\pm0.08$ & $1.64\pm0.02$ &$0.20\pm0.01$               & $2.21_{-0.02}^{+0.02}$ &UVM2	& $1.34\pm0.03$ &0.918 & $185/1.37$  \\ \hline


    S79 (09/05/23)	    &   $47.98\pm4.39$   &$7.34\pm2.40$ &$5.50\pm0.84$ & $0.23\pm 0.06$    &$2.82_{-0.28}^{+0.32}$ &UVW2	& $2.17\pm0.12$ &$0.883$ & $61.88/0.73$\\
    S80 (13/05/23)	    &   $40.79\pm4.18$   &$11.79\pm3.99$ &$4.91\pm0.88$ & $0.13\pm 0.04$    &$2.42_{-0.61}^{+0.50}$ &UVW2	& $2.02\pm0.11$ &$0.890$ & $70.32/1.00$\\
    S81 (17/05/23)	    &   $47.25\pm4.48$   &$8.18\pm2.75$ &$5.23\pm0.72$ & $0.20\pm 0.04$    &$2.65_{-0.37}^{+0.29}$ &UVW2	& $2.29\pm0.11$ &$0.901$ & $52.35/0.77$\\
    S82 (${21/05/23}$)	    &   $40.30\pm3.86$   &$5.09\pm2.05$ &$3.63\pm0.88$ & $0.19\pm 0.03$    &$2.67_{-0.54}^{+0.38}$ &UVW2	& $1.96\pm0.11$ &$0.936$ & $73.22/1.05$\\
    S83 (${20/05/23}$)	    &   $32.15\pm2.40$   &$8.74\pm1.85$ &$4.03\pm0.38$ & $0.17\pm 0.03$    &$2.35_{-0.27}^{+0.23}$ &UVW2	& $1.82\pm0.12$ &$0.934$ & $76.65/0.72$\\
    S84 (22/05/23)	    &   $47.78\pm4.43$   &$6.00\pm1.95$ &$4.81\pm0.66$ & $0.22\pm 0.05$    &$2.90_{-0.30}^{+0.30}$ &UVW2	& $2.05\pm0.11$ &$0.896$ & $92.00/1.15$\\
    S86 (23/05/23)	    &   $49.47\pm7.17$   &$5.25\pm2.47$ &$4.26\pm0.82$ & $0.20\pm 0.06$    &$2.96_{-0.75}^{+0.64}$ &UVW2	& $2.20\pm0.16$ &$0.928$ & $27.88/0.59$\\
    S87 (25/05/23)	    &   $51.18\pm5.12$   &$9.86\pm3.13$ &$5.63\pm0.76$ & $0.18\pm 0.05$    &$2.64_{-0.40}^{+0.29}$ &UVW2	& $1.81\pm0.13$ &$0.849$ & $50.49/0.78$\\
    S88 (26/05/23)	    &   $48.63\pm6.83$   &$5.02\pm2.04$ &$5.75\pm1.48$ & $0.29\pm 0.08$    &$3.14_{-0.47}^{+0.46}$ &UVW2	& $1.76\pm0.13$ &$0.841$ & $57.01/1.24$\\
    S89 (09/06/23)	    &   $45.36\pm2.95$   &$7.43\pm1.95$ &$4.21\pm0.42$ & $0.18\pm 0.02$    &$2.45_{-0.38}^{+0.28}$ &UVW2	& $1.96\pm0.11$ &$0.911$ & $100.76/0.87$\\
    S90 (11/06/23)	    &   $44.17\pm4.22$   &$21.93\pm11.75$ &$4.42\pm0.79$ & $0.16\pm 0.02$    &$1.92_{-0.49}^{+0.39}$ &UVW2	& $2.05\pm0.11$ &$0.910$ & $48.38/0.78$\\
    S91 (12/06/23)	    &   $45.29\pm4.62$   &$8.07\pm3.94$ &$4.30\pm0.77$ & $0.17\pm 0.03$    &$2.46_{-0.69}^{+0.58}$ &UVW2	& $2.05\pm0.11$ &$0.912$ & $47.20/0.76$\\
    S92 (13/06/23)	    &   $50.97\pm4.68$   &$9.54\pm3.01$ &$5.03\pm0.68$ & $0.17\pm 0.03$    &$2.49_{-0.46}^{+0.32}$ &UVW2	& $2.02\pm0.11$ &$0.886$ & $67.84/0.92$\\
    S94 (15/06/23)	    &   $45.28\pm2.71$   &$8.60\pm1.79$ &$4.58\pm0.40$ & $0.17\pm 0.02$    &$2.53_{-0.27}^{+0.21}$ &UVW2	& $2.00\pm0.11$ &$0.900$ & $113.26/0.83$\\
    S95 (17/06/23)	    &   $47.88\pm2.82$   &$7.30\pm1.70$ &$4.26\pm0.38$ & $0.18\pm 0.02$    &$2.46_{-0.32}^{+0.25}$ &UVW2	& $1.94\pm0.12$ &$0.907$ & $100.74/0.79$\\
    S96 (21/06/23)	    &   $48.16\pm4.10$   &$5.37\pm1.64$ &$4.06\pm0.52$ & $0.19\pm 0.04$    &$2.95_{-0.27}^{+0.25}$ &UVW2	& $2.17\pm0.13$ &$0.934$ & $80.38/0.90$\\
    S97 (24/06/23)	    &   $43.41\pm3.64$   &$6.99\pm2.217$ &$4.42\pm0.54$ & $0.19\pm 0.04$    &$2.70_{-0.33}^{+0.26}$ &UVW2	& $1.92\pm0.11$ &$0.899$ & $55.14/0.63$\\
    S98 (27/06/23)	    &   $52.34\pm4.77$   &$5.81\pm1.91$ &$5.33\pm0.87$ & $0.23\pm 0.05$    &$2.93_{-0.31}^{+0.33}$ &UVW2	& $1.96\pm0.11$ &$0.872$ & $63.01/0.77$\\
    S100 (03/07/23)	    &   $47.17\pm3.96$   &$5.25\pm1.73$ &$3.93\pm0.55$ & $0.19\pm 0.04$    &$2.91_{-0.31}^{+0.28}$ &UVW2	& $2.15\pm0.13$ &$0.938$ & $55.61/0.67$\\
    S101 (06/07/23)	    &   $47.61\pm3.86$   &$9.38\pm3.72$ &$3.91\pm0.56$ & $0.17\pm 0.01$    &$2.07_{-0.68}^{+0.54}$ &UVW2	& $2.05\pm0.11$ &$0.931$ & $70.60/0.89$\\
    S102 (09/07/23)	    &   $44.35\pm3.49$   &$8.59\pm3.22$ &$4.20\pm0.62$ & $0.17\pm 0.02$    &$2.38_{-0.67}^{+0.40}$ &UVW2	& $2.09\pm0.11$ &$0.922$ & $75.05/0.83$\\
    S103 (15/07/23)	    &   $47.51\pm4.03$   &$7.85\pm2.39$ &$4.90\pm0.62$ & $0.19\pm 0.03$    &$2.61_{-0.35}^{+0.28}$ &UVW2	& $2.29\pm0.13$ &$0.911$ & $77.34/0.94$\\
    S104 (25/07/23)	    &   $47.61\pm3.86$   &$9.38\pm3.72$ &$3.91\pm0.56$ & $0.17\pm 0.10$    &$2.69_{-0.59}^{+0.32}$ &UVW2	& $2.26\pm0.18$ &$0.905$ & $38.51/0.77$\\
    S105 (27/07/23)	    &   $46.23\pm5.70$   &$14.58\pm9.48$ &$5.51\pm1.15$ & $0.18\pm 0.03$    &$1.91_{-0.68}^{+0.76}$ &UVW2	& $2.07\pm0.14$ &$0.875$ & $42.19/0.98$\\
    S106 (02/08/23)	    &   $60.17\pm5.24$   &$7.85\pm2.50$ &$6.19\pm0.87$ & $0.21\pm 0.04$    &$2.80_{-0.30}^{+0.28}$ &UVW2	& $2.09\pm0.14$ &$0.857$ & $65.07/0.79$\\
    S107 (05/08/23)	    &   $41.35\pm3.39$   &$8.83\pm2.53$ &$4.00\pm0.49$ & $0.16\pm 0.02$    &$2.31_{-0.40}^{+0.33}$ &UVW2	& $2.11\pm0.15$ &$0.932$ & $87.54/0.97$\\
    S108 (08/08/23)	    &   $49.76\pm4.70$   &$8.04\pm2.85$ &$5.29\pm0.83$ & $0.20\pm 0.07$    &$2.72_{-0.36}^{+0.33}$ &UVW2	& $1.91\pm0.12$ &$0.868$ & $56.44/0.81$\\
    S109 (11/08/23)	    &   $50.12\pm3.95$   &$6.19\pm2.18$ &$4.09\pm0.54$ & $0.18\pm 0.03$    &$2.70_{-0.42}^{+0.31}$ &UVW2	& $1.89\pm0.09$$^{*}$ &$0.910$ & $68.30/0.77$\\
    S110 (14/08/23)	    &   $60.79\pm5.63$   &$7.12\pm2.58$ &$5.43\pm0.76$ & $0.20\pm 0.05$    &$2.96_{-0.33}^{+0.32}$ &UVW2	& $2.07\pm0.14$$^{*}$ &$0.892$ & $60.05/0.77$\\
    S111 (17/08/23)	    &   $50.15\pm4.08$   &$9.45\pm3.52$ &$4.51\pm0.65$ & $0.17\pm 0.02$    &$2.32_{-0.63}^{+0.42}$ &UVW2	& $2.02\pm0.11$$^{*}$ &$0.904$ & $69.34/0.83$\\
    S113 (26/08/23)	    &   $44.67\pm3.92$   &$7.74\pm2.76$ &$4.38\pm0.62$ & $0.17\pm 0.03$    &$2.59_{-0.47}^{+0.34}$ &UVW2	& $1.91\pm0.12$$^{*}$ &$0.900$ & $83.81/1.03$\\
    S115 (23/09/23)	    &   $56.49\pm4.74$   &$9.78\pm2.81$ &$5.21\pm0.72$ & $0.18\pm 0.05$    &$2.27_{-0.55}^{+0.39}$ &UVW2	& $1.68\pm0.09$$^{*}$ &$0.845$ & $49.81/0.66$\\
    S116 (27/09/23)	    &   $54.52\pm4.73$   &$12.06\pm3.97$ &$6.39\pm0.83$ & $0.18\pm 0.05$    &$2.52_{-0.40}^{+0.30}$ &UVW2	& $1.59\pm0.04$$^{*}$ &$0.806$ & $72.23/0.90$\\
    S117 (01/10/23)	    &   $46.85\pm2.98$   &$9.96\pm2.49$ &$5.12\pm0.50$ & $0.19\pm 0.02$    &$2.18_{-0.41}^{+0.30}$ &UVW2	& $2.31\pm0.15$$^{*}$ &$0.906$ & $92.11/0.77$\\
    S118 (05/10/23)	    &   $44.67\pm3.92$   &$7.74\pm2.76$ &$4.38\pm0.62$ & $0.19\pm 0.05$    &$2.56_{-0.32}^{+0.27}$ &UVW2	& $1.17\pm0.09$$^{*}$ &$0.769$ & $80.79/0.94$\\
    S119 (09/10/23)	    &   $58.67\pm5.48$   &$4.13\pm2.16$ &$5.00\pm0.91$ & $0.22\pm 0.03$    &$3.09_{-0.41}^{+0.48}$ &UVW2	& $1.74\pm0.10$$^{*}$ &$0.862$ & $69.78/0.88$\\
    S120 (13/10/23)	    &   $58.64\pm4.37$   &$8.14\pm2.32$ &$6.45\pm0.85$ & $0.22\pm 0.04$    &$2.78_{-0.27}^{+0.26}$ &UVW2	& $--$ &$--$ & $78.63/0.76$\\
    S121 (17/10/23)	    &   $53.42\pm4.20$   &$10.73\pm3.81$ &$5.75\pm0.71$ & $0.19\pm 0.03$    &$2.40_{-0.57}^{+0.35}$ &UVW2	& $1.57\pm0.10$$^{*}$ &$0.822$ & $80.18/0.86$\\
    S122 (21/10/23)	    &   $54.63\pm4.47$   &$9.54\pm3.31$ &$5.88\pm0.72$ & $0.19\pm 0.06$    &$2.69_{-0.37}^{+0.31}$ &UVW2	& $1.55\pm0.10$$^{*}$ &$0.816$ & $87.44/0.96$\\
    

\hline 
\end{tabular}  
$^{\rm A}$ Flux in units of $ 10^{-12}\funit{}$ corrected for Galactic absorption.\\
$^{\rm B}$ UV flux density in units of $ 10^{-15}\funita{}$ and $\alpha_{\rm OX}=-0.385\log(\rm F_{2\kev}/F_{2500\rm \AA})$\\
The UV flux density was corrected for Galactic absorption using the correction magnitude of $\rm A_{\lambda}=0.690$ obtained from NED.\\

\end{table*}

\begin{table*}
\fontsize{8}{10}
\centering
  \caption{The spectral parameters obtained using {\it Swift} UV and X-ray observations of 1ES~1927+654.  \label{Table:swift_obs2}} 
  \begin{tabular}{cccccccccccc} \hline\hline 

ID(DD/MM/YY)			&$F_{0.3-2\kev}^{\rm A}$	&$F_{2-10\kev}^{\rm A}$ &$F_{1.5-2.5\kev}^{\rm A}$  & kT (keV) & $\Gamma$ & UV filter& UV flux density$^{\rm B}$ &$\alpha_{\rm OX}$ & $\rm \chi^2/ \chi^2_{\nu}$ \\ \hline  

X1 (20/05/11)	    &   $9.41\pm0.66$   & $3.92\pm0.08$ & $1.64\pm0.02$ &$0.20\pm0.01$               & $2.21_{-0.02}^{+0.02}$ &UVM2	& $1.34\pm0.03$ &0.918 & $185/1.37$  \\ \hline


S123 (24/10/23)	    &   $52.45\pm4.47$   &$9.92\pm2.85$ &$5.67\pm0.67$ & $0.18\pm 0.04$    &$2.58_{-0.37}^{+0.28}$ &UVW2	& $1.68\pm0.10$$^{*}$ &$0.835$ & $71.72/0.82$\\
S124 (06/11/23)	    &   $49.06\pm6.18$   &$10.97\pm5.32$ &$5.21\pm0.93$ & $0.17\pm 0.04$    &$2.32_{-0.77}^{+0.54}$ &UVW2	& $1.74\pm0.10$$^{*}$ &$0.855$ & $26.66/0.67$\\
S125 (10/11/23)	    &   $45.21\pm4.36$   &$11.53\pm4.00$ &$5.88\pm0.80$ & $0.19\pm 0.05$    &$2.40_{-0.47}^{+0.32}$ &UVW2	& $2.35\pm0.13$ &$0.885$ & $66.76/0.95$\\
S126 (15/11/23)	    &   $53.16\pm4.58$   &$7.40\pm3.29$ &$4.16\pm0.64$ & $0.17\pm 0.02$    &$2.37_{-0.72}^{+0.56}$ &UVW2	& $--$ &$--$ & $64.64/0.82$\\
S127 (18/11/23)	    &   $47.85\pm4.05$   &$12.62\pm3.73$ &$5.72\pm0.68$ & $0.18\pm 0.03$    &$2.28_{-0.45}^{+0.33}$ &UVW2	& $1.65\pm0.10$$^{*}$ &$0.831$ & $90.42/1.06$\\
S128 (22/11/23)	    &   $58.52\pm4.39$   &$10.88\pm2.86$ &$5.75\pm0.70$ & $0.16\pm 0.03$    &$2.57_{-0.36}^{+0.27}$ &UVW2	& $1.72\pm0.09$$^{*}$ &$0.837$ & $68.23/0.69$\\
S129 (26/11/23)	    &   $54.03\pm5.47$   &$9.52\pm3.52$ &$5.08\pm0.77$ & $0.17\pm 0.02$    &$2.45_{-0.51}^{+0.35}$ &UVW2	& $1.39\pm0.08$$^{*}$ &$0.822$ & $49.59/0.86$\\
S130 (30/11/23)	    &   $47.57\pm4.40$   &$13.43\pm5.64$ &$5.33\pm0.78$ & $0.18\pm 0.02$    &$1.97_{-0.59}^{+0.54}$ &UVW2	& $2.26\pm0.13$$^{*}$ &$0.895$ & $61.67/0.87$\\
S131 (04/12/23)	    &   $54.79\pm3.34$   &$7.85\pm1.80$ &$5.72\pm0.56$ & $0.21\pm 0.03$    &$2.71_{-0.19}^{+0.22}$ &UVW2	& $1.57\pm0.11$$^{*}$ &$0.823$ & $116.26/0.89$\\
S132 (08/12/23)	    &   $53.56\pm4.22$   &$12.08\pm3.01$ &$6.15\pm0.66$ & $0.18\pm 0.02$    &$2.38_{-0.35}^{+0.26}$ &UVW2	& $1.59\pm0.10$$^{*}$ &$0.813$ & $95.73/0.85$\\
S133 (13/12/23)	    &   $59.10\pm5.57$   &$9.34\pm4.00$ &$5.75\pm0.71$ & $0.17\pm 0.02$    &$2.46_{-0.53}^{+0.38}$ &UVW2	& $1.54\pm0.10$$^{*}$ &$0.819$ & $48.32/0.76$\\
S134 (25/12/23)	    &   $56.98\pm4.01$   &$7.81\pm2.00$ &$5.52\pm0.59$ & $0.20\pm 0.03$    &$2.74_{-0.24}^{+0.21}$ &UVW2	& $1.72\pm0.10$$^{*}$ &$0.844$ & $100.93/0.91$\\
S135 (28/12/23)	    &   $51.02\pm4.14$   &$8.10\pm2.89$ &$5.08\pm0.69$ & $0.20\pm 0.02$    &$2.35_{-0.58}^{+0.37}$ &UVW2	& $1.62\pm0.09$$^{*}$ &$0.848$ & $76.28/0.90$\\
S137 (05/01/24)	    &   $55.71\pm4.86$   &$8.77\pm2.80$ &$6.81\pm1.07$ & $0.22\pm 0.04$    &$2.64_{-0.38}^{+0.31}$ &UVW2	& $1.15\pm0.09$$^{*}$ &$0.742$ & $68.78/0.83$\\
S138 (09/01/24)	    &   $54.95\pm4.91$   &$7.72\pm2.80$ &$5.45\pm0.76$ & $0.20\pm 0.03$    &$2.62_{-0.50}^{+0.35}$ &UVW2	& $1.60\pm0.09$$^{*}$ &$0.834$ & $62.49/0.81$\\
S139 (13/01/24)	    &   $54.46\pm4.16$   &$10.68\pm3.89$ &$5.88\pm0.68$ & $0.18\pm 0.03$    &$2.50_{-0.39}^{+0.28}$ &UVW2	& $1.31\pm0.09$$^{*}$ &$0.788$ & $90.08/0.93$\\
S140 (17/01/24)	    &   $64.76\pm5.17$   &$9.72\pm3.03$ &$6.66\pm0.81$ & $0.20\pm 0.05$    &$2.78_{-0.30}^{+0.26}$ &UVW2	& $1.24\pm0.09$$^{*}$ &$0.758$ & $80.64/0.82$\\
S140A (24/01/24)	    &   $52.36\pm5.00$   &$9.31\pm3.09$ &$5.80\pm0.82$ & $0.20\pm 0.03$    &$2.40_{-0.33}^{+0.48}$ &UVW2	& $1.91\pm0.11$ &$0.853$ & $75.16/1.03$\\
S140B (03/02/24)	    &   $52.97\pm4.33$   &$16.78\pm7.31$ &$6.50\pm0.92$ & $0.20\pm 0.03$    &$1.94_{-0.87}^{+0.67}$ &UVW2	& $1.68\pm0.11$ &$0.813$ & $63.6/0.79$\\
S140C (21/02/24)	    &   $61.15\pm4.97$   &$9.69\pm3.00$ &$6.66\pm0.81$ & $0.19\pm 0.04$    &$2.70_{-0.35}^{+0.27}$ &UVW2	& $1.91\pm0.10$ &$0.830$ & $83.13/0.98$\\
S141 (28/02/24)	    &   $62.76\pm7.15$   &$7.52\pm3.64$ &$5.63\pm1.04$ & $0.19\pm 0.04$    &$2.78_{-0.57}^{+0.44}$ &UVW2	& $2.00\pm0.11$ &$0.866$ & $40.88/0.83$\\
S141A (10/03/24)	    &   $63.39\pm7.15$   &$9.78\pm2.61$ &$6.51\pm0.68$ & $0.20\pm 0.03$    &$2.66_{-0.28}^{+0.23}$ &UVW2	& $1.81\pm0.13$ &$0.855$ & $90.03/0.77$\\
S143 (06/04/24)	    &   $74.7\pm4.96$   &$9.71\pm2.67$ &$6.08\pm0.78$ & $0.20\pm 0.05$    &$2.81_{-0.30}^{+0.24}$ &UVW2	& $2.15\pm0.13$ &$0.865$ & $89.82/0.92$\\
S144A (11/04/24)	    &   $54.71\pm5.20$   &$13.79\pm5.82$ &$5.18\pm0.75$ & $0.17\pm 0.02$    &$1.93_{-0.96}^{+0.76}$ &UVW2	& $2.02\pm0.11$ &$0.881$ & $64.06/0.79$\\


\hline 
\end{tabular}  
$^{\rm A}$ Flux in units of $ 10^{-12}\funit{}$ corrected for Galactic absorption.\\
$^{\rm B}$ UV flux density in units of $ 10^{-15}\funita{}$ and $\alpha_{\rm OX}=-0.385\log(\rm F_{2\kev}/F_{2500\rm \AA})$\\
The UV flux density was corrected for Galactic absorption using the correction magnitude of $\rm A_{\lambda}=0.690$ obtained from NED.\\

\end{table*}

\begin{table*}
\centering
  \caption{X-ray spectral parameters as obtained from fitting the NuSTAR observations, using standard powerlaw model.}\label{Table:nustar_table}
  \begin{tabular}{cccccccc} \hline\hline 

Models	&  Parameter(units) & May 2023 & June 2023 & August 2023 & September 2023 \\
\hline
 Gal. abs. & $\nh(\times) 10^{20}\cmsqi$ & 6.42 (frozen) &  --- & --- & --- \\
 Intrinsic. abs. & $\nh(\times) 10^{20}\cmsqi$ & 6.00(frozen) &  --- & --- & --- \\
power-law   &   $\Gamma$                & $3.21^{+0.07}_{-0.07}$ & $3.21^{+0.07}_{-0.07}$ & $3.11^{+0.07}_{-0.07}$ & $3.20^{+0.07}_{-0.07}$ \\
            &    norm($10^{-2}$)        & $1.43^{+0.19}_{-0.17}$ & $1.48^{+0.18}_{-0.16}$ & $1.23^{+0.15}_{-0.14}$ & $1.52^{+0.21}_{-0.18}$\\

\hline
\hline
$\rm \chi^2/dof$ &                         &   &$519/562$  \\
\hline
\end{tabular} 
\end{table*}

\section{Discussion}\label{sec:discussion}
 In this paper, we report an episode of large-amplitude (factor of $\sim 8$) monotonic increase of the soft X-ray $0.3-2\kev$ in the CL-AGN 1ES~1927+654, over roughly the same timescale as the GHz radio emission increased by a factor of 40-60 as reported in the companion paper by \cite{meyer2024}. Since Aug 2023 till Apr-2024 the core-radio emission at 5 GHz has plateaued. However, since Feb-2024 \cite{meyer2024} detected a nascent jet at $0.1\pc$ scale which has progressed to $0.4\pc$ by Apr 2024 at a speed of $0.2$c, marking an extraordinary discovery of a jet forming and evolving in real time in a changing-look AGN. Our second companion paper, Masterson et al. 2025, detected a mHz QPO in May-2022 ($\nu=0.93\pm 0.06$ mHz) which have been consistently detected till March 2024 with an increasing frequency ($\nu=2.35\pm 0.05$ mHz as on March 2024). For the first time in this source a QPO has been consistently detected for $\sim 2$ years.
 
 
 Below we recap the most important observational results from our extensive multi-wavelength campaign: 
 
 \begin{itemize}
     
\item The soft-excess is still increasing in flux and is now $\sim 10$ times its pre-changing-look value (May-2011). The overall soft X-ray ($0.3-2\kev$) flux is $\sim 8$ times higher than the pre-changing-look phase. The best fit black-body model temperature (describing the soft X-ray excess)  is very well constrained in a narrow range $0.13-0.16\kev$. We find that the total integrated energy (May 2022- Apr 2024) under the \swift{}-XRT and UVOT light curves in the $0.3-2\kev$, the $2-10\kev$ and the UV bands are $1.56\times 10^{51}$ erg, $3.12\times 10^{50}$ erg, and $5.76\times 10^{50}$ erg, respectively. The UV accretion energy is clearly lower than the energy pumped into the X-rays.

\item The $2-10\kev$ power law slope became softer, $\Gamma=2.70\pm 0.04$ in May 2022 to $\Gamma=3.27\pm 0.04$ in March 2024. With \nustar{} $3-40\kev$ spectra we measured a cut-off temperature of the X-ray corona ($kT_{\rm e}=8^{+8}_{-2}\kev$). 

\item Coincident with the radio-flare in Feb-Aug 2023, the \xmm{} and \nustar{} observations interestingly show no change in the power law slope or the $2-10\kev$ flux ($<5\%$). 

\item We do not detect significant optical or UV flux variability (only $<30\%$), and only $<2$ times variation in the $2-10\kev$ flux in this epoch of study (May 2022- Apr-2024).

\item We detect no weeks-month time scale correlation in the variability between the three quantities: UV flux-density, soft X-ray excess flux and $2-10\kev$ flux. However, in the longer term ($\sim 2$ years) we detect a trend of increasing spectral slope ($\Gamma$) with increasing $2-10\kev$ flux, and also an increasing $2-10\kev$ flux with the soft X-ray excess flux. Though we note that the rise in the $2-10\kev$ flux is $\lesssim 2$ times.

\item The ratio of the 5 GHz to $2-10\kev$ flux in this source was $L_{\rm 5\, GHz}/L_{2-10\kev} \sim 10^{-5.5}$ in May 2022 indicating a radio emission dominated by corona in this radio-quiet source, which is now moving towards the jet dominated ratio of $\sim 10^{-3}$ (See Fig \ref{fig:radio_xray_corr})

\item We do not detect any FeK$\alpha$ emission line, or any absorption in the $0.3-10\kev$ spectra. 

\end{itemize}

  In light of these observational results, we discuss the following scientific topics in this section.  We use a black hole mass of $\sim 10^6\msol$ \citep{Li_broad_line_1es}, which gives a gravitational radius of $r_g\equiv GM/c^2=10^{11}~{\rm cm}$.



\begin{figure*}
    \centering
     \includegraphics[width=5.9cm]{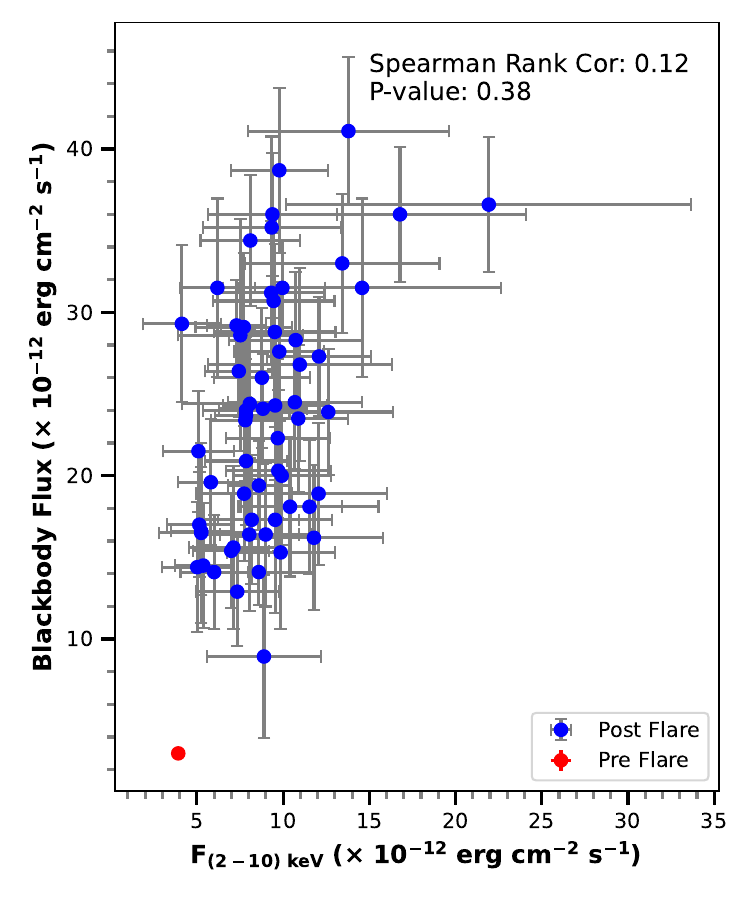}
     \includegraphics[width=5.9cm]{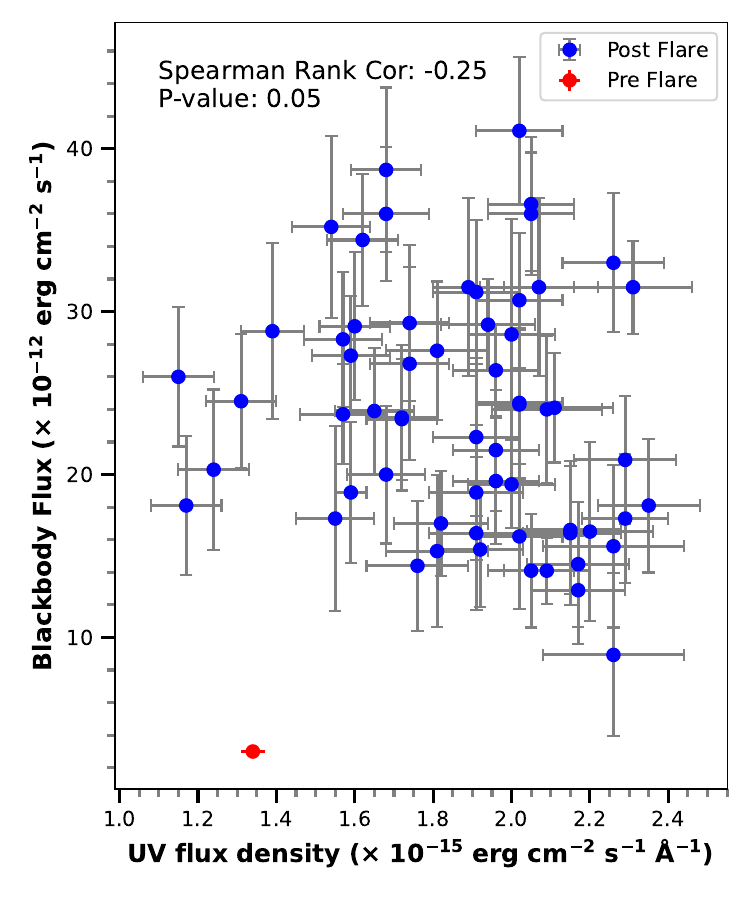}
     \includegraphics[width=5.9cm]{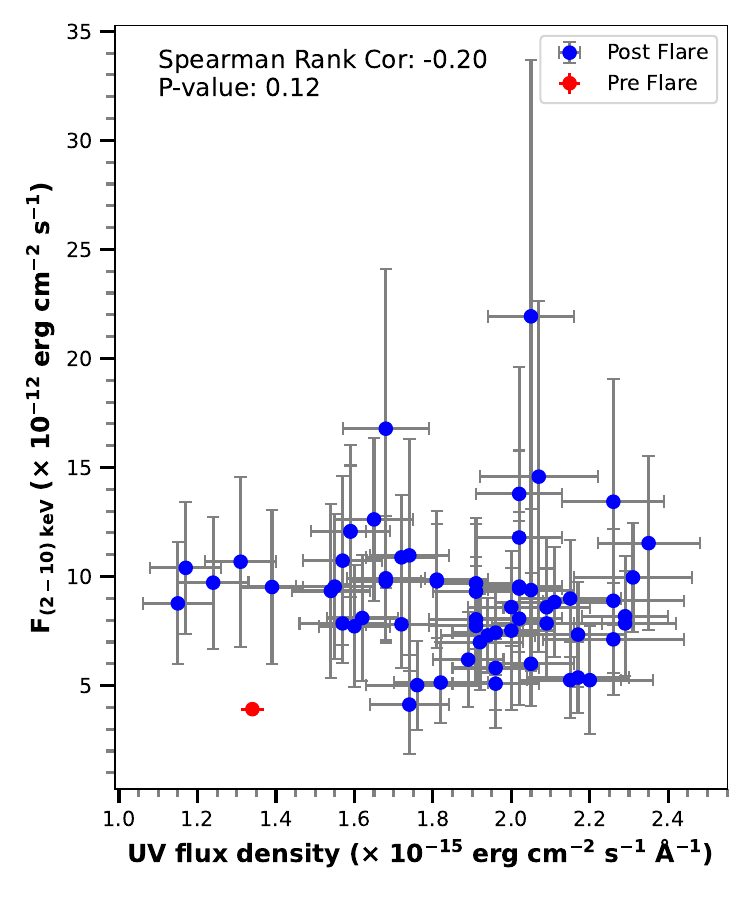}

    \caption{ The correlations between the UV flux density, the black body flux (soft-excess) and the power law $2-10\kev$ flux as obtained from \swift{} weekly-monthly cadence observations in the time period May 2022- April 2024. The red dot denote the pre-flare May 2011 observation, and is not used in the correlation calculations. {\it Left panel:} The blackbody flux vs the hard X-ray flux ($2-10\kev $) , {\it Middle panel:} The blackbody vs the UV flux density, and the {\it Right panel:} $2-10\kev$ flux vs the UV flux density. The Spearman rank correlation coefficient and the p-values are quoted in the panels. None of them show any statistically significant correlation.} 
    \label{fig:swift_correlations}
\end{figure*}

\begin{figure*}[]
 \hbox{
\includegraphics[height=6cm, width=9cm] {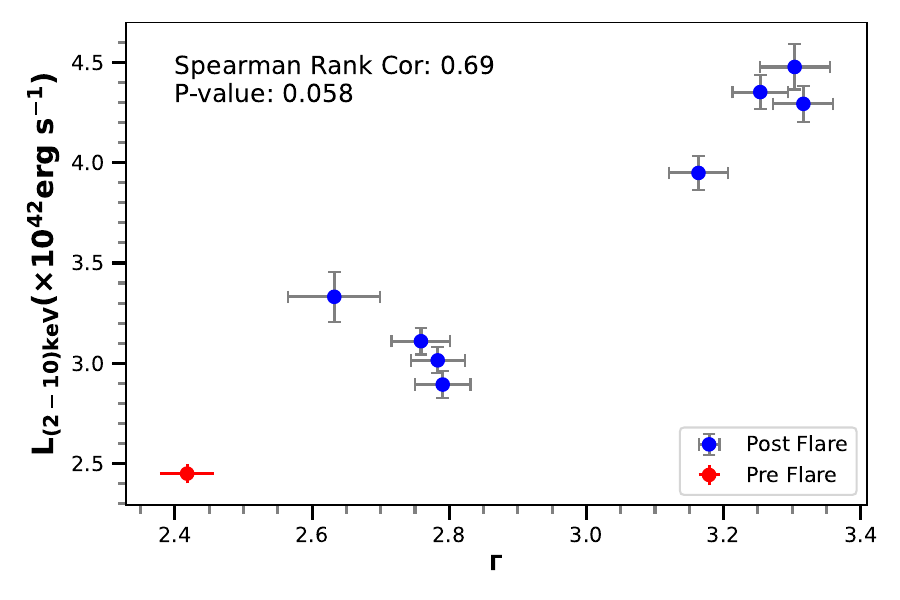}
\includegraphics[height=6cm, width=9cm]{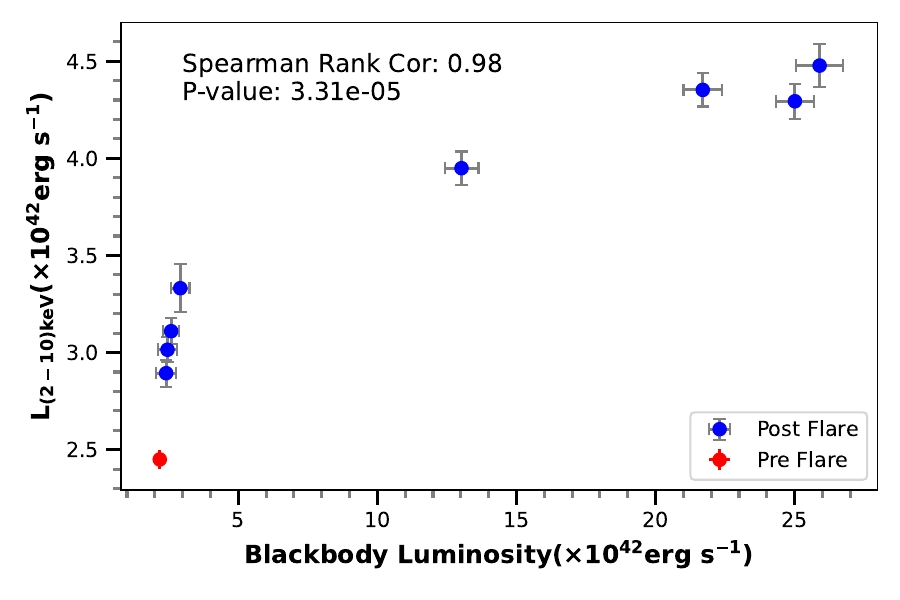}
}
\caption{The correlations between the black body (soft-excess) luminosity, power law slope $\Gamma$, and the power law $2-10\kev$  luminosity as obtained from \xmm{} observations in the time period July 2022- March 2024.  {\it Left:} The correlation between the coronal $2-10\kev$ luminosity and the power law slope $\Gamma$ . {\it Right:} The correlation between the power law $2-10\kev$ luminosity and the black-body luminosity. Table \ref{Table:xmm_pnfit} lists the parameters. }
\label{fig:xmm_correlations}
\end{figure*}

\subsection{The jet launching mechanism}\label{ssec:energetics}

 Our high-cadence, multi-wavelength observations capture the launching of a jet in real time, presenting a unique opportunity to probe jet launching mechanisms.
Here we discuss two prominent jet launching mechanisms: the Blandford-Znajek (BZ;~\citet{bz77}) mechanism, which extracts rotational energy from the black hole itself, and the Blandford-Payne (BP;~\citet{blandfordpayne}) mechanism, which extracts rotational energy from the accretion disk. 
The arguments for/against the BZ and BP models can be summarized into four categories: 
i) outflow collimation, ii) outflow speed, iii) outflow trigger, and iv) outflow efficiency.

\begin{figure*}[]
 \centering
{\includegraphics[width=18cm, height=7cm]{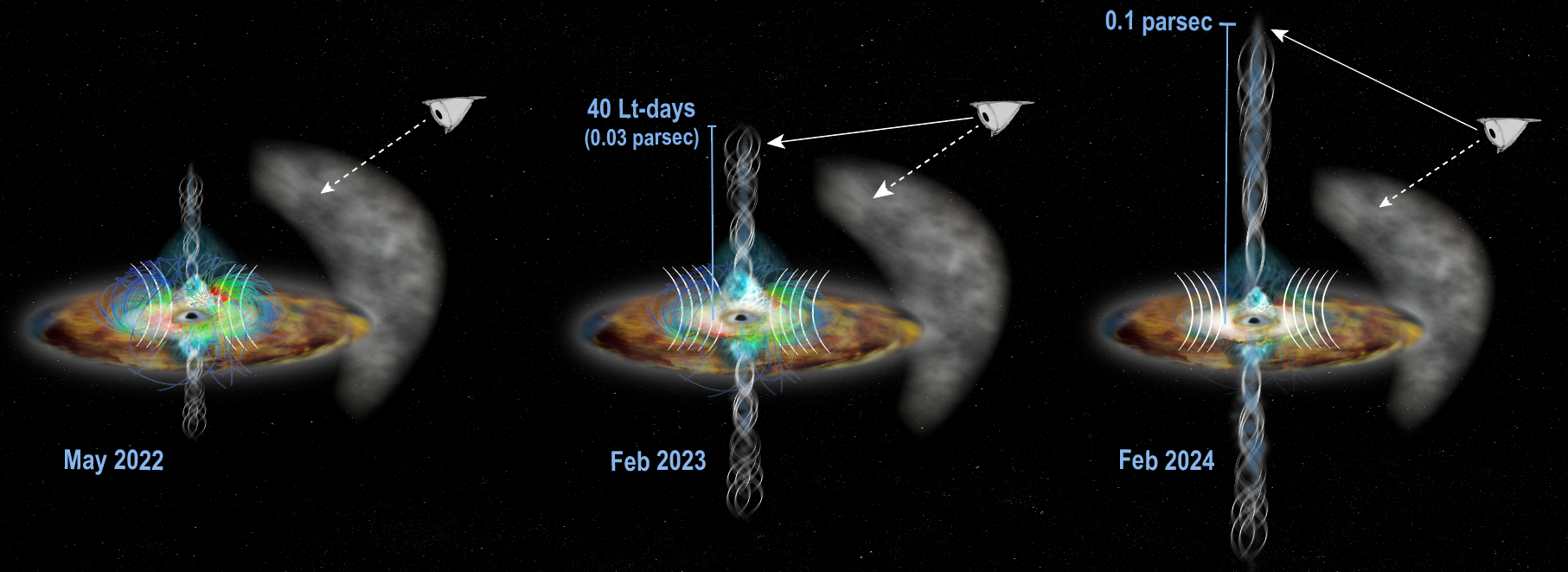}}
\caption{ Cartoon depicting the spectacular accretion-disk, corona and jet evolution in the enigmatic changing-look AGN 1ES 1927+654. From left to right gives us the temporal evolution of the system which includes the (a) soft X-ray rise, (b) QPO detection, (c) jet formation and (d) jet evolution. {\it Left panel (May 2022):} This is the time when the soft X-ray started to rise, and the jet and the QPO ($\sim 0.91$ mHz) were formed. The accretion disk is shown in brown, the random magnetic field are denoted by colored lines (red and cyan), the advected poloidal fields (responsible for jet formation) are shown in white. The tenuous white cloud corresponds to the external screen of hot gas that blocks the expected radio emission via free-free absorption. This hot screening gas may possibly be the broad emission line (BLR) clouds located at $\sim 30-40$ light days \citep[as detected by][]{trak19}. The eye denotes the observer and the dotted lines denote the line of sight to the central engine, while the solid lines denote a view towards the jet.  We note that although the jet may be formed at the same time as the soft X-ray rise and the QPO, we could not detect it because of the free-free absorption of the screening gas. {\it Middle panel (Feb 2023):} Same as the left panel, but the jet has now evolved beyond $30-40$ light -days ($0.03\pc$) and hence has come out of the BLR screen, which is visible as the radio flare continuing for a few months (Feb-Aug 2023). The radio emission gets plateaued once the jet has fully come out (sometime in Aug 2023). The soft X-rays from the inner disk (white patch in the disk) has gained in strength compared to that in May 2022. Variability in the \xmm{} soft X-ray lightcurve on the order of 200-300s constrains the soft X-ray emission to originate from within $\sim100r_g$ of the black hole. {\it Right panel (Feb 2024):}  Same as the middle panel, but the jet has now evolved beyond $0.1\pc$, which corresponds to the spatial resolution of {\it VLBA}, and hence we can detect spatially resolved bi-polar jets. The soft X-rays are now extremely bright (by a factor of $\sim 8$) compared to that in May 2022. In all the panels note that the UV disk (denoted in brown) has a near-steady rate of accretion as also the corona (the cyan triangular shape at the center) which does not show much variability in flux. See Tables \ref{Table:xmm_pnfit}, \ref{Table:nustar_table} and Figure \ref{fig:xray_uv_alpha_ox_zoomed} for reference}.
\label{fig:cartoon}

\end{figure*}

\subsubsection{Outflow Collimation}
The collimation of the 1ES 1927 outflow suggests a BZ ``jet'' origin rather than a BP disk ``wind'' origin.
A disk wind generally has a wide, nearly isotropic opening angle, compared to a much narrower, more collimated relativistic jet.
Therefore, the spatially-resolved, bi-lobed structure of the radio emission reported by~\citet{meyer2024}, supports the BZ mechanism over the BP mechanism.

\subsubsection{Outflow Speed}
Radio measurements suggest a mildly relativistic outflow speed of $0.2c$~\citep{meyer2024}.
BP winds can have speeds of $0.2c$ only if they are launched from quite close ($\lesssim10r_g$) to the black hole~\citep{blandfordpayne}, while a speed of $0.2c$ is somewhat low for BZ mechanism~\citep{bz77}.
However, this low speed could be an inclination angle effect.
If the jet contains a faster-moving jet spine that is Doppler beamed away from the line-of-sight, the observations may only measure the slower-moving jet sheath. 
Future observations of the jet's evolution will better constrain its speed and therefore help discriminate between jet-launching mechanisms. 

\subsubsection{Outflow trigger}
The proposed trigger for the outflow is a reconfiguration of the magnetic field close to the black hole.
The magnetic energy density close to the black hole likely remains relatively constant since the hard X-ray flux, powered by reconnection/turbulence in the corona increases by less than a factor of two over the observational period.
The flux needed to change the dominant magnetic components close to the black hole could have accumulated since the previous 1ES~1927 CL event associated with an inversion of the magnetic field polarity in a magnetically-dominated accretion disk~\citep{scepi21, laha2022}. 
Since that CL event, accretion through a canonical thin disk~\citep{Shakura_Sunyaev_1973} onto the black hole proceeds normally, as inferred from the relatively constant UV flux (Figure \ref{fig:xray_uv_alpha_ox_zoomed} panel 4). 
As accretion occurs, magnetic flux can advect onto the black hole over the viscous timescale $t_{\rm visc}$:
\begin{equation}
t_{visc} = 18\left(\frac{r}{100 r_g}\right)^{3/2} \left(\frac{M}{10^6 M_{\odot}}\right) \alpha_{0.1}^{-1} h_{0.1}^{-2}~{\rm months}
\end{equation}
where $r$ is the radial distance from the black hole, $M$ is the black hole mass, $\alpha_{0.1}$ is the viscosity parameter in units of 0.1, and $h_{0.1}$ is the disk aspect ratio $H/r$ in units of 0.1. 
For these estimated upper limit parameters, the 18-month interval between the CL event and the start of the radio flare would imply that the magnetic flux was advected from a spatial scale of $100r_g$.
This spatial scale decreases for smaller values of $\alpha$ and $h$. 
However, it will increase if the inner region is a thick flow rather than a thin disk~\citep{dexter2019}.

If poloidal magnetic flux is advected towards the event horizon from the inner accretion disk, it could trigger either the BZ or BP mechanism since both rely on the configuration of the magnetic field~\citep{blandfordpayne, bz77}.
In principle, this advection could occur without changing the accretion rate, and therefore maintain a relatively constant UV flux. 
However, a BP wind would likely remove angular momentum from the disk~\citep{blandfordpayne}, thereby lowering the accretion rate and decreasing the UV flux from the accretion disk.
Therefore, the near-steady UV flux suggests that the jet launch occurs due to the BZ mechanism.

\subsubsection{Outflow Efficiency}
The radio power of the 1ES 1927 outflow implies a high efficiency.
The power expected for a BZ jet from a rapidly-spinning (i.e. maximally efficient) black hole with spin parameter $a\approx1$ is

\begin{equation}
    P_{\rm BZ} \approx\frac{\kappa\pi c}{4}r_g^2B^2 \, \, \, {\rm erg/s},~\label{eq:Pbz}
\end{equation}
where $\kappa \simeq 0.05$ ~\citep{tchekhovskoy2011} and B is the net vertical magnetic flux accumulated on the black hole horizon, i.e. at length scales of a few $r_g\sim10^{-7}$ pc.

The magnetic field close to the black hole is likely well above equipartition values due to the nature of the corona.
The corona comprises electrons with temperatures $\sim10^9$ K that sit a distance $R_c\lesssim10r_g$ from the black hole \citep[X-ray variability and microlensing constraints;][]{fabian+09,dai+10,kara+13,wilkins+21,laha_coronareview_arxiv_2024}.
These electrons inverse Compton scatter disk photons to produce hard X-rays~\citep[e.g.][]{katz76,pozdnyakov+77}.
To maintain these high electron temperatures, the corona must be magnetically-dominated~\citep{merloni2001}.
Assuming that some fraction of the magnetic energy in the X-ray corona converts into hard X-rays yields a {\it minimum} coronal magnetic field strength of \begin{equation}
    B_0\ge10^4~{\rm G}\left(\frac{L_{2-10}}{10^{43}~{\rm erg/s}}\right)^{1/2}\left(\frac{R_c}{10r_g}\right)^{-1}\left(\frac{M}{10^6M_\odot}\right)^{-1},
\end{equation}
where the 1ES 1927 luminosity $L_{2-10}\sim 10^{43}~{\rm erg/s}$ (Table \ref{Table:xmm_pnfit}). 
Recent analytic calculations and particle-in-cell simulations of the dissipative and radiative processes in the coronal plasma yield magnetic fields of up to
$B_0\sim10^8~{\rm G}(M/10M_\odot)^{-1/2}=3\times10^5~{\rm G}$~\citep[e.g.][]{beloborodov2017,groselj+24}.
Large $B_0$ values on event horizon scales are feasible considering that measurements from AGN jets have previously found magnetic field strengths of $\sim0.1$ G on $\sim1$ pc scales from core frequency-shift methods~\citep{osullivan2009} and $\sim10$ G on $\sim0.1$ pc scales from Faraday rotation measurements~\citep{martividal2015}.
Recent very long baseline interferometry measurements have also suggested kilo-Gauss fields close to the black hole~\citep{lisakov2024}.
Such observational values are consistent with $B_0 \gtrsim 10^5$~G at the base of the jet with a $1/r$ decay of the magnetic field, and are thereby consistent with theoretical and numerical predictions for launching relativistic jets~\citep{bz77, tchekhovskoy2011}.

Plugging this value for the magnetic field into Eq.~\ref{eq:Pbz}, the BZ power is $P_{\rm BZ}\approx 3\times10^{42}~{\rm erg/s}$.
This power is on the order of the jet kinetic power estimated from the elevated radio state of 1ES~1927+654~\citep{meyer2024}. 
The high jet kinetic power implies that the BZ jet must convert into radiation at close to 100\% efficiency. 
Because the BZ mechanism is more efficient than the BP mechanism, this high efficiency suggests that the jet launches via the BZ mechanism.



\subsection{Characteristics of the radio emission}

\subsubsection{The radio emission as optically-thin synchrotron emission}
The observed radio spectral slopes between 5--8.4 GHz and 8.4--23.6 GHz indicate a nature consistent with a small-scale synchrotron jet/outflow \citep{meyer2024}. The spectrum is curved and peaks around 5 GHz and resembles the typical profile of Gigahertz-peaked spectrum (GPS) AGN sources. The relatively steep late-time spectral index ($\alpha$ where $F_\nu\sim \nu^{-\alpha}$) between 8.4 and 23.6 GHz ($\alpha=1.05\pm 0.26$) suggests that the emission is dominated by synchrotron processes from an optically thin region, aligning with the characteristics expected from jet emissions.
 For the expected magnetic field of $\sim0.2$ G at 0.1 pc, an electron with Lorentz factor $\gamma=100$ emits synchrotron emission at 5 GHz.
 Such an electron would lose half its energy due to synchrotron emission after travelling for about a parsec, i.e. at 0.1 pc it has not lost much energy due to synchrotron cooling (the slow cooling regime).
In the slow cooling regime, the spectral index $\alpha$ relates to the underlying electron distribution's power-law index $p$ as $p=2\alpha+1$ rather than $p=2\alpha-1$ as in the fast cooling regime~\citep{blumenthal1970}.
Therefore, the measured 5 - 8 GHz emission's $\alpha\sim0.5$ \citep{meyer2024} gives an electron power-law index of $p\sim 2$, which is a reasonable value from magnetic reconnection.
The increase in the 8 - 22 GHz band to $\alpha\sim1$ suggests the presence of a cooling break in that frequency range, indicating that the electrons now emit most of their energy as synchrotron radiation.

\subsubsection{The 200 day delay of the radio flare from the start of the soft X-ray rise}\label{ssec:delay_discussion}
The delay in the radio increase of $\sim200$ days relative to the soft X-ray rise could be due to obscuration by an external screen of hot gas that blocks the expected emission via free-free absorption until the jet emerges from behind it (See Figure \ref{fig:cartoon}). 
If this external screen sits at the estimated distance of the BLR, approximately 30-40 light-days \citep[$\sim 10^{17}$ cm~][]{trak19}, the screen will absorb all synchrotron emission until the jet propagates past it, i.e. after $\sim200$ days for a jet speed of $0.2c$ \citep{meyer2024}.
The frequency where an ionized gas at a temperature of $10^4$ K becomes optically thin to free-free absorption is proportional to $nL^{1/2}$, where $n$ is the number density of the gas and $L$ is the path length through the absorbing gas. 
The necessary combination of number density and path length could come from a localized overdensity from e.g. gas that was expelled during the previous CL event or compressed due to radiation pressure~\citep{baskin2021}.
We note that the length of the delay requires that the free-free absorption comes from an external screen rather than from hot gas spatially co-located with the jet. 
The higher frequency X band $\sim 8$ GHz became optically thin before the lower frequency $\sim 5$ GHz during the radio flare~\citep{meyer2024}, which is consistent with the optical depth from free-free absorption $\tau\sim \nu^{-2}$, though not exclusive to it. Figure \ref{fig:cartoon}
 shows the formation and evolution of jet, along with the soft X-ray rise in this source. The effect of the external screen is also depicted in the cartoon. In addition, if the radio emission were from synchrotron self-absorbed electrons only on smaller scales, say $\sim1000 r_g$, the radio rise should occur after a propagation time of only $\approx 2$ days for the measured jet velocity of $0.2c$. Hence the delay must be due to an external screen.


\begin{figure}[h!]
 \centering
{\includegraphics[height=8cm, width=9cm] {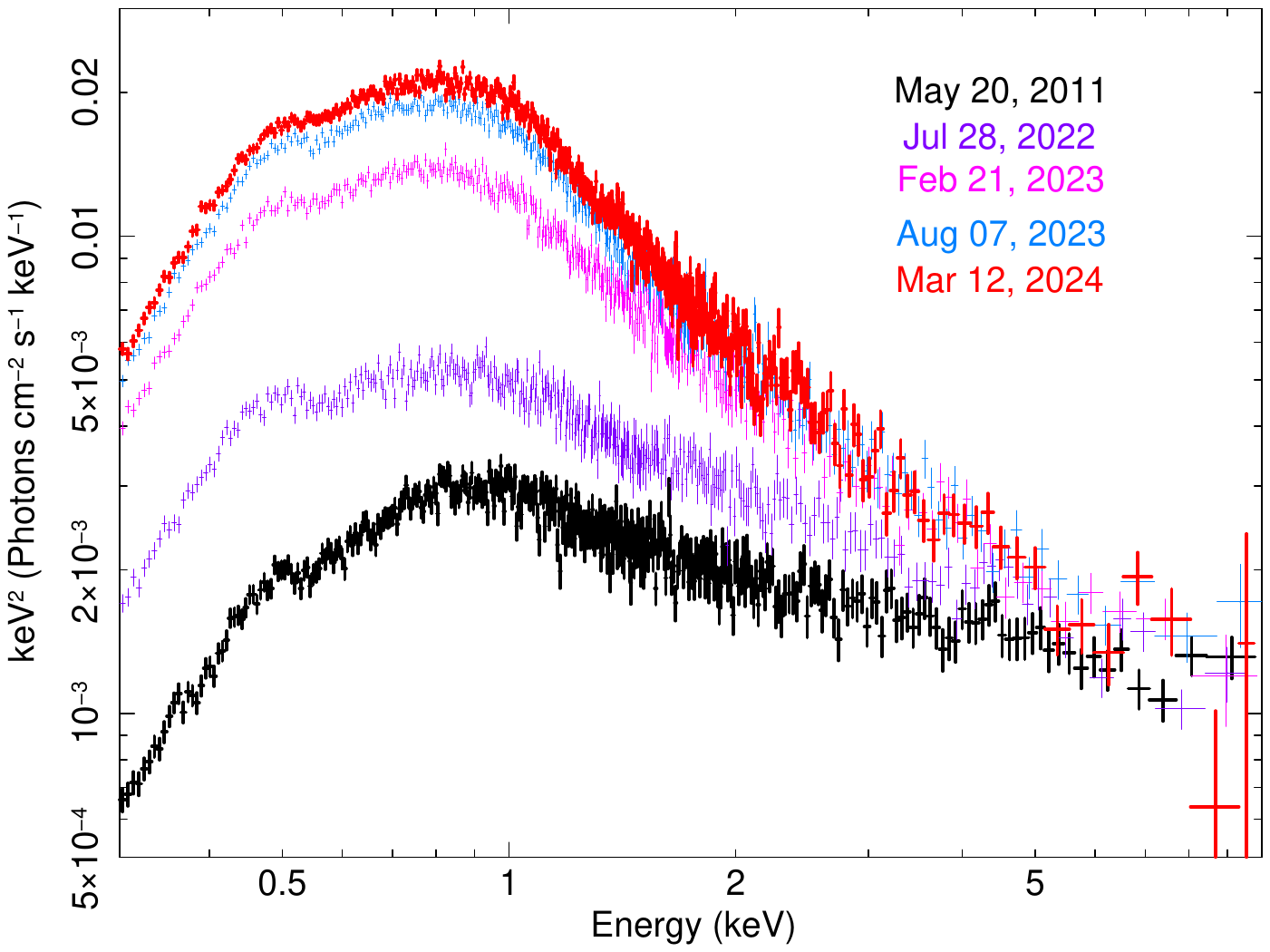}}
\caption{The over-plotted EPIC-pn spectra of 1ES~1927+654 showing the evolution of the total energy dissipated per $\kev$ across the $0.3-10\kev$ band. The different colors denote the different dates of observation. The topmost spectrum (in red) with the highest energy in the soft X-rays denote the latest observation in March 2024. In comparison, the pre-flare (or pre-changing-look) phase in May 2011 in black show much less energy in the soft X-ray band. We note here that the $2-10\kev$ band does not show radical changes for different epochs, compared to the soft X-rays. We have only included observations from July 2022- March 2024 to cover the soft X-ray rise, and plotted the May 2011 observation for comparison. See Table \ref{Table:xmm_pnfit} for details. The figure has been created using the {\tt setplot area} and {\tt plot uuef} commands in {\it XSPEC}.}
\label{fig:pn_overplot}
\end{figure}

\subsection{The nature of the soft-excess} \label{ssec:softXray-discussion}

The origin of soft X-ray-excess in RQ-AGN is highly debated \citep[see e.g.][]{crummy2006,Vaia_softexcess_2024}. In 1ES~1927+654 the evolution/rise of the soft-excess (modeled by black body) coincided with the formation and evolution of the radio jet, along with the softening of the power law emission and an appearance of a QPO. Overall, we note that: (1) The soft-excess light curve monotonically increased by a factor of $\sim 10$ (Table \ref{Table:xmm_pnfit}), (2) The temperature of the black body is very narrowly constrained in $0.14-0.16\kev$ during the entire rise phase (See Tables \ref{Table:xmm_pnfit}, \ref{Table:swift_obs1} and \ref{Table:swift_obs2}), and there is no correlation between the soft-excess flux and the temperature (3) the soft X-ray, hard X-ray and UV fluxes do \emph{not} correlate with each other, indicating that they arise from physically distinct regions.  
This new soft-excess can neither be explained with i) the hard X-ray disk-reflection model \citep{Garcia2014} because we observe no substantial variation in the $2-10\kev$ flux, nor with ii) the `warm Comptonization' scenario \citep{Done2012} because we do not observe any increase in the UV flux in a similar time-span \footnote{This model could, however, be applicable if the assumption that the energy source for the soft-excess is the standard disk accretion (in UV) is relaxed.}. All these point to the fact that the origin of the new soft excess in 1ES~1957+654 is unique and not the one we find in RQ-AGN. The origin of the soft X-ray-excess can possibly be related with the available magnetic energy and the jet emission.

 The total integrated energy under the \swift{}-XRT and UVOT light curves in the $0.3-2\kev$, the $2-10\kev$ and the UV bands are $1.56\times 10^{51}$ erg, $3.12\times 10^{50}$ erg, and $5.76\times 10^{50}$ erg, respectively. The total X-ray energy surpasses that of the UV, indicating that a source of energy-extraction other than that of the standard accretion disk (UV) is at play. This is also demonstrated in Figure \ref{fig:pn_overplot} where the soft X-rays show a larger energy dissipation compared to the hard X-rays. Additionally, the total jet power and soft excess luminosity are comparable ($\sim 10^{43}\lunit$), suggesting that the rise of soft X-rays could be magnetically powered, for example through a change in the magnetic field topology or the presence of the jet. For example, the increase in soft X-rays after the jet launches could be related to a change in magnetic field structure in the innermost disk, or a change in the emitting volume caused by the presence of the jet. Variability in the \xmm{} soft X-ray lightcurve on the order of 200-300s constrains the soft X-ray emission to originate from within $\sim100r_g$ of the black hole. Detailed spectral and timing analysis of the soft X-ray spectra will be carried out in a future work.

\begin{figure}[h!]
\centering
\includegraphics[trim=0mm 0cm 0mm 0cm, clip, scale=0.30]{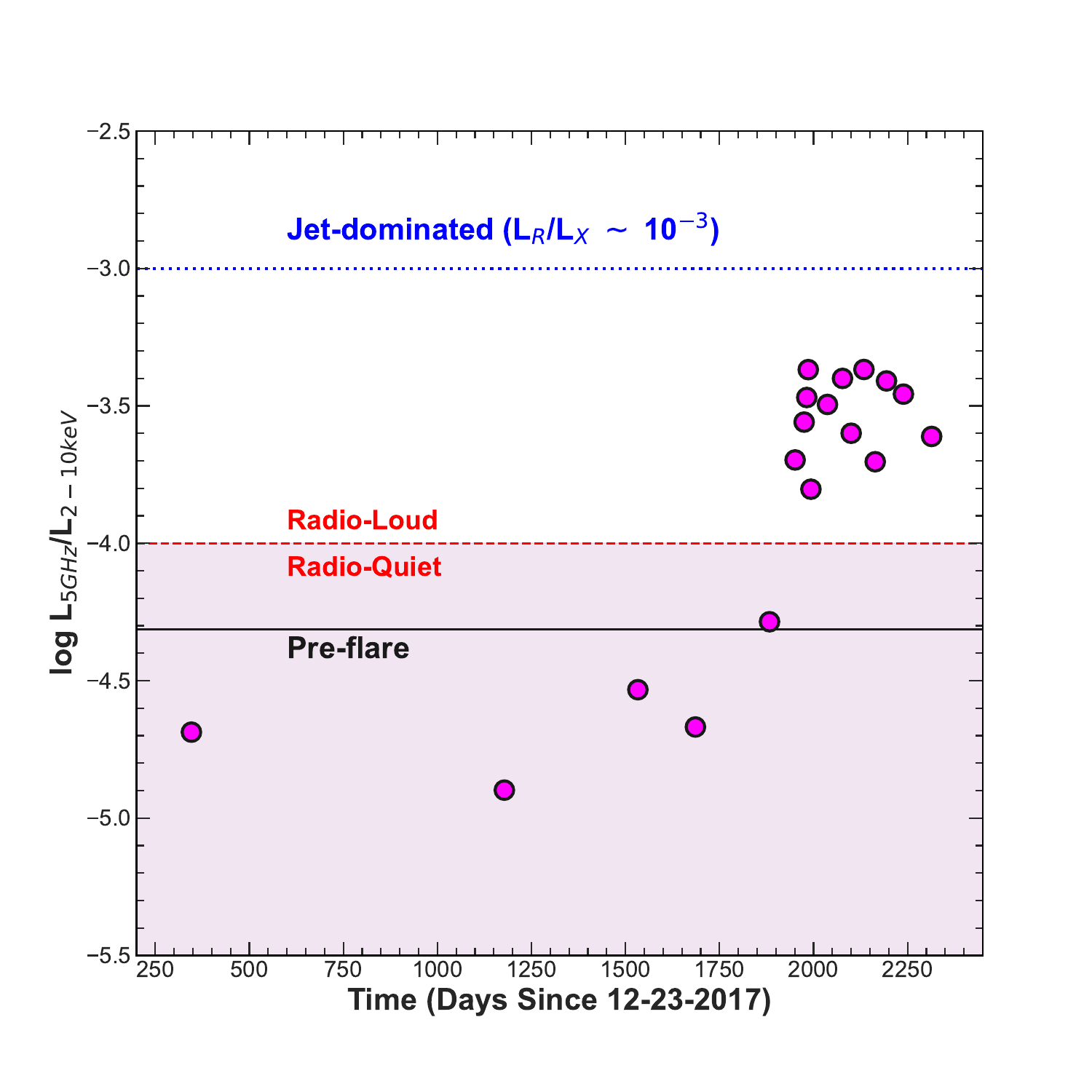}
\caption{\textbf{1ES~1927+654 jumping out of the radio-quiet regime}. The ratio between 5 GHz radio peak luminosities and 2-10 keV X-ray luminosities (aka radio loudness parameter) are plotted as a function of days since the flare on Dec. 23, 2017. The black solid line represents the preflare ratio, with semicontemporaneous radio and X-ray fluxes taken from~\cite{laha2022}. The blue dotted line marks the boundary between radio-loud (RL) and radio-quiet (RQ) AGNs as introduced by \citet{panessa2007}, distinguishing Seyferts from low luminosity radio galaxies (LLRGs) of $L_\mathrm{R}$/$L_\mathrm{X}$ $\approx$ 10$^{-3}$. The red dashed line denotes the classical division between RL and RQ AGNs by~\citet{Terashima_2003ApJ...583..145T}.
The colored-filled area indicates the phase-space (10$^{-6}$< $L_\mathrm{R}$/$L_\mathrm{X}$ < 10$^{-4}$) encompassed by coronal emission from radio-quiet AGN, based on the well-established relation for coronally active cool stars of~$L_\mathrm{R}$/$L_\mathrm{X}$ = 10$^{-5}$~\citep[][]{laor08,guedel93}{}.}

             \label{fig:radio_xray_corr}
\end{figure}

\subsection{The unusual coronal emission and the disk-corona-jet relation}

The $2-10\kev$ coronal emission in the source 1ES~1927+654 has been unusual since the CL event. In Appendix Figure \ref{fig:xray_uv_alpha_ox_whole} we show the entire multi-wavelength light curve of this source since the previous CL phenomenon (Dec 2017) till April 2024. This is perhaps the only source where the corona completely vanished and reappeared in a month timescale during the CL event \citep{Ricci_2020,Ricci2021,laha2022}. The pre-CL power-law slope ($\Gamma=2.45$) has been on the higher end of AGN $\Gamma$ distribution \citep{tortosa2018}. During the violent CL event just prior to the vanishing of the corona in Aug 2018 we detected a very soft power law slope $\Gamma\sim 4-5$ \citep{laha2022,Ricci_2020,Ricci2021,ricci_NatAs_CLAGNreview_2023}, very unusual for an AGN, along with an abnormally low electron temperature ($kT_{\rm e}\sim 1.5 \kev$). As the X-ray corona was destroyed and then recreated the spectral slope was again very soft with $\Gamma\sim3$ \citep{Ricci_2020}.

Very recently in May 2022 after the soft X-ray started to rise along with the advent of a QPO and a radio-jet, we noticed that the coronal slope again gradually became softer (over a period of $\sim 1$ year), from $\Gamma= 2.79$ in Aug 2022 to $\Gamma= 3.32$ in March 2024. With broadband \nustar{} observations in May-Sept 2023 coinciding with the exponential radio rise, we measured a constant coronal slope of $\Gamma \sim 3.2$, and an electron temperature of $kT_{\rm e}=8.4^{+8.8}_{-2.6}\kev$ (with the thermal Comptonization model {\tt nthcomp} in {\it XSPEC}). {The corona is cooler than the typical ones found in normal AGNs. } Interestingly, low coronal temperatures have been observed in super-Eddington systems \citep{tortosa2023} and predicted theoretically from GRRMHD simulations \citep{Pacucci_Narayan_2024} and analytical calculations \citep{Madau_2024}. However, 1ES~1927+654 is still sub-Eddington as obtained in Section \ref{subsubsec:swift_reprocess} ($\lambda_{Edd} \approx 0.3$, and hence it is unusual. 


The ratio of the 5 GHz and $2-10\kev$ flux in AGN have been known to show an interesting correlation also exhibited in coronally active stars \citep{guedel93, laor08}, known as the Gudel-Benz (GB) relation, with a typical value of $10^{-5.5}$. During the violent changing look event in 2018-2019 till the end of 2022, the radio to X-ray flux ratio of 1ES~1927+654 was within the GB range \citep{laha2022, ghosh2023}, indicating that the bulk of radio emission was from the corona. The 5 GHz core ($<1\pc$ scale) radio emission was lowest during the coronal destruction in 2018 indicating again that the low level of radio emission ($\sim 2 $ mJy) had a coronal origin. Since early 2023 when the radio flux started to rise, the ratio assumed higher values, mostly because there no associated increase in the $2-10\kev$ flux. We note that the hard X-ray $3-40\kev$ flux and its slope, as determined from four \nustar{} observations coinciding with the radio exponential rise, remain unchanged at the $<5\%$ level (Figure \ref{fig:overplot_nustar_xmm} and Table \ref{Table:nustar_table}). This indicated that the excess radio emission is coming from something other than the corona, possibly a jet. In Figure~\ref{fig:radio_xray_corr}, we have plotted the evolution of the GB ratio of the source during the radio flare and we note that the initially radio-quiet source is now moving towards the jet dominated ratio of $\sim 10^{-3}$ \citep{panessa2007}.

\subsection{The QPO and the jet relation}\label{ssec:qpo_discussion}

The rise of the soft X-ray flux in 2022 coincides with the emergence of a QPO feature in the X-rays (Masterson et al. 2025), which could be when the jet is formed as well. The QPO is more strongly detected in the hard X-ray band (above $2\kev$), with much less significance in the soft X-ray band possibly indicating that the QPO is primarily associated with the ``corona" of the AGN.  Very few SMBHs have shown QPOs to date, such as RE~J1034+396 \citep{gierlinski2008} and ASASSN-14li \citep{pasham2019}. To date, no SMBH QPO shows the dramatic frequency evolution that is seen in the QPO in 1ES 1927+654 (see Masterson et al. 2025 for further comparisons with existing SMBH QPOs).

Observations have indicated that in black hole binaries (BHBs), the advent of QPOs is sometimes also associated with a radio outburst and a subsequent ejection of a radio jet. Type-B QPOs are closely related to the production of jet-like outflows \citep{corbel2001,gallo2004,miller-jones2012,russell2019}, but a physical model is yet to be developed. As a BHB transitions from the hard to a soft state, a strong (type-C) low-frequency QPO appears in the power spectral density \citep{remillard2002,casella2005}. Their frequencies increase from $\sim 0.01$ to $10$ Hz as the spectrum softens. After some time during this transition, the type-C QPOs and the associated strong band-limited red noise are replaced by a type-B QPO ($\sim 4-9$ Hz) with a considerably weaker band-limited noise \citep{casella2005,belloni2010,ingram2019,homan2020}, which is then associated with a radio outburst and a jet. This has never been observed in an AGN. The new QPO in 1ES 1927 is remarkable, in that (1) it is consistently found for $\sim 2$ years, (2) its frequency increases with time, (3) the QPO is most prevalent in the hard X-rays, and (4) and its frequency ($\nu=1-2$ mHz) does not seem to line up with a BHB analogy, because a simple linear scaling of the frequency with mass to a Type-B QPO would require the BH mass of 1ES~1927+654 to be $<5\times 10^4\msol$). The time-evolution of its frequency (increase in frequency with time) also does not follow the BHB pattern. However, the near-simultaneous occurrence of the QPO and the jet in 1ES~1927+654 points to physics very closely related to the ones in BHB, and is an important discovery in the field of AGN. For example, the increase of the frequency indicates that the physical size of the QPO emitting region is shrinking with time. This is in agreement with the evolution of the X-ray spectral index, indicating more intense Compton cooling as expected in a more compact corona. Similar behaviour is routinely seen in X-ray binaries, whereby all characteristic frequencies increase as the spectrum softens \citep{belloni2010}. This is often interpreted in the context of the truncated disk model \citep{done2007}, whereby the disk inner radius and therefore the size of the corona reduces as the spectrum softens. Reflection spectroscopy reveals that the disk inner radius does indeed reduce as the luminosity increases, even though there is widespread disagreement over the precise location of the disk inner radius \citep[see e.g., ][]{garcia2015}. Suggested physical mechanisms for a moving truncation radius include evaporation/condensation \citep{eardley1975,mayer2007} and magnetic truncation \citep{liska2022}. Our companion work (Masterson et al. 2025) explores a variety of explanations for the QPO in 1ES~1927+654 and we refer the reader to that paper for further details. A future work will carry out a detailed theoretical study connecting the jet and the QPO of this source, and also explore other possible models/interpretations.

\section{Conclusions}\label{sec:conclusions}
 In this paper, we report an episode of large-amplitude (factor of $\sim 8$) monotonic increase of the soft X-ray $0.3-2\kev$ in the CL-AGN 1ES~1927+654, over roughly the same timescale as the GHz radio emission increased by a factor of 40-60 as reported in the companion paper by \cite{meyer2024}. In addition, \citep{meyer2024} detected a spatially resolved radio jet evolving at a speed of $\sim 0.2c$ in $0.1-0.3\pc$ scale, and Masterson et al. 2025 detected a consistent QPO in the hard X-rays with increasing frequency, both of which are extraordinary and rare events in an AGN. We list below the most important conclusions from this extensive multi-wavelength study.

\begin{itemize}

  \item Jet emission mechanism: The weak variation of the $2-10\kev$ X-ray emission and the near-steady UV
emission suggest that the magnetic energy density and accretion rate are
relatively unchanged, and that the jet could be launched due to a
reconfiguration of the magnetic field (toroidal to poloidal) close to
the black hole. Advecting poloidal flux onto the event horizon would
trigger the Blandford-Znajek (BZ) mechanism, leading to the onset of the
jet.  The concurrent softening of the coronal slope (from $\Gamma=
2.70\pm 0.04$ to $\Gamma=3.27\pm 0.04$), occurrence of a QPO, and low coronal temperature
($kT_{e}=8_{-3}^{+8}\kev$) during the radio outburst suggest that the poloidal
field reconfiguration can significantly impact coronal properties and
thus influence jet dynamics. These extraordinary findings in real
time are crucial for coronal and jet plasma studies, particularly as
our results are independent of coronal geometry.

      \item The absorbing screen and late time jet evolution: A possible explanation for the 200 days delay of the radio flare from the start of the soft X-ray rise, could be due to a screen of hot gas that blocks the expected emission via free-free absorption until the jet emerges from behind it. If this external screen sits at the estimated distance of the BLR, approximately 30-40 light-days \citep[$\sim 10^{17}$ cm~][]{trak19}, the screen will absorb all synchrotron emission until the jet propagates past it, i.e. after $\sim200$ days for a jet speed of $0.2c$. See Figure \ref{fig:cartoon}.

    \item The origin of the soft X-ray excess: We note that: (1) The soft-excess light curve  monotonically increased by a factor of $\sim 10$ (2) The temperature of the black body (modeling the soft-excess) is very narrowly constrained in $0.14-0.16\kev$ during the entire rise phase, (3) variability in the \xmm{} soft X-ray lightcurve on the order of 200-300s constrains the soft X-ray emission to originate from within $\sim100r_g$ of the black hole, and (4) there is no correlation between the soft-excess flux with the hard X-ray and UV fluxes. The total integrated energy under the \swift{}-XRT and UVOT light curves in the $0.3-2\kev$, the $2-10\kev$ and the UV bands are $1.56\times 10^{51}$ erg, $3.12\times 10^{50}$ erg, and $5.76\times 10^{50}$ erg, respectively. The energetics of the UV accretion therefore cannot account for the jet and the soft X-ray rise, implying an additional source of energy in action. A gradual rise in the soft excess by a factor of $\sim 10$ times in 2 years, along with the radio-flare, and their mutually similar luminosity $L_{\rm soft-excess}\sim L_{\rm radio-jet}\sim 10^{43}\lunit$, points towards a related origin, which could be powered by the magnetic fields in the inner regions of the accretion disk.

      \item The Jet-QPO relation: The occurrence of a QPO near-simultaneously with the jet signals some interesting similarity with black hole binaries where observations have detected QPO and jets to occur simultaneously. However, there is no direct analog of the QPO frequency and its evolution as detected in 1ES~1927+654 with the BHB systems. As the QPO frequency increased over a time period of 2 years, we detected a steeper photon-index ($\Gamma\sim 3.3$) and a cool corona ($kT_{\rm e}=8.4^{+8.8}_{-2.6}\kev$), accompanied by the formation and evolution of the jet. We refer to Masterson et al., 2025 for several other interpretation of the QPO origin.

      \item The Jet-corona relation: We do not find any connection between the radio and the $2-10\kev$ (coronal) flux variations over the period of two years (May 2022- April 2024). The $2-10\kev$ flux varied only by a factor $<2$, with a softening power law slope ($\Gamma=2.7-3.2$), while the radio emission flared by a factor of $\sim 60$ and plateaued at a high flux state. In particular, we do not detect any change ($<5\%$) in coronal flux and power law slope $\Gamma$ during the exponential radio flare (Feb- Aug 2023). The ratio of the 5 GHz to $2-10\kev$ flux in this source was $L_{\rm 5\, GHz}/L_{2-10\kev} \sim 10^{-5.5}$ in May 2022 indicating a radio emission dominated by corona in this radio-quiet source, which is now moving towards the jet dominated ratio of $\sim 10^{-3}$ (See Figure \ref{fig:radio_xray_corr}).

\end{itemize}



\section{Acknowledgements}

The material is based upon work supported by NASA under award number 80GSFC21M0002. 
JS acknowledges the Czech Science Foundation project No.22-22643S. SL is thankful to Jay Friedlander who has created the graphics of the jet and the heated gas (Fig \ref{fig:cartoon}). Claudio Ricci acknowledges support from Fondecyt Regular grant 1230345 and ANID BASAL project FB210003.
FR and SB acknowledge funding from PRIN MUR 2022 SEAWIND 2022Y2T94C, supported by European Union - Next Generation EU. SB acknowledges support from INAF LG 2023 BLOSSOM. Main Pal thanks the support of the Inter-University Centre for Astronomy and Astrophysics (IUCAA), Pune through the Visiting Associate Programme.
Matt Nicholl is supported by the European Research Council (ERC) under the European Union’s Horizon 2020 research and innovation programme (grant agreement No.~948381) and by UK Space Agency Grant No.~ST/Y000692/1. SL, EM and EB acknowledge
support from NSF-BSF grant numbers: NSF-2407801,
BSF-2023752.



\newpage
\appendix  

\begin{figure*}[h!]
    \centering
    \includegraphics[height=21cm,width=17cm]{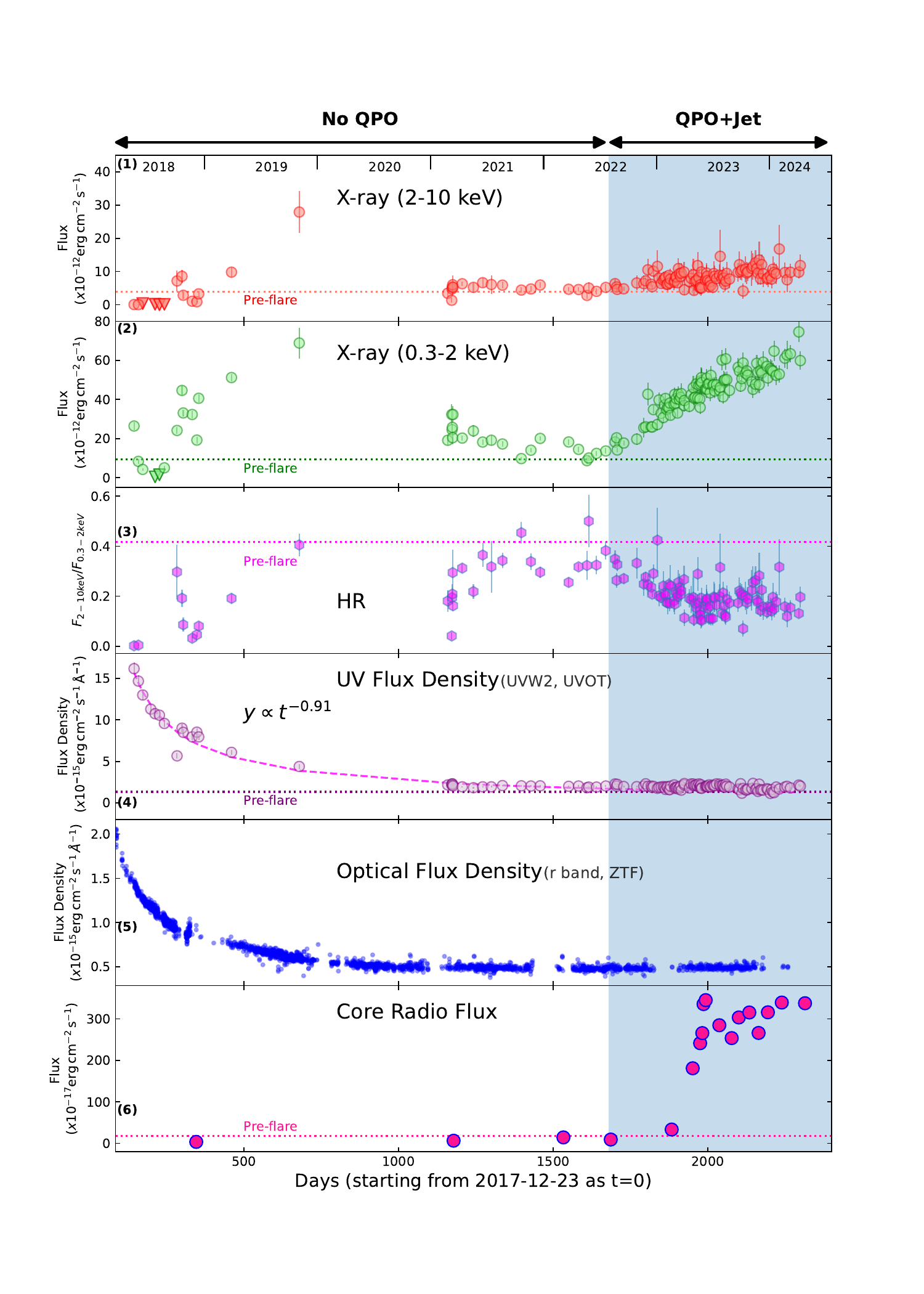}
    \caption{The X-ray, UV, optical, and radio light curves of 1ES1927+654 during the period May 2018 (when the first CL outburst happened, and \swift{} monitoring started) to April 2024. The recent QPO+Jet+Soft-X-ray rise phase is shaded in blue. See Table \ref{Table:swift_obs1} and \cite{laha2022,ghosh2023} for details. The different panels are the same as the ones discussed in Figure \ref{fig:xray_uv_alpha_ox_zoomed}.    
    } 
    \label{fig:xray_uv_alpha_ox_whole}
\end{figure*}

\begin{figure*}[h!]
 \centering

\vbox{
 
 \hbox{
    \includegraphics[width=7.9cm]{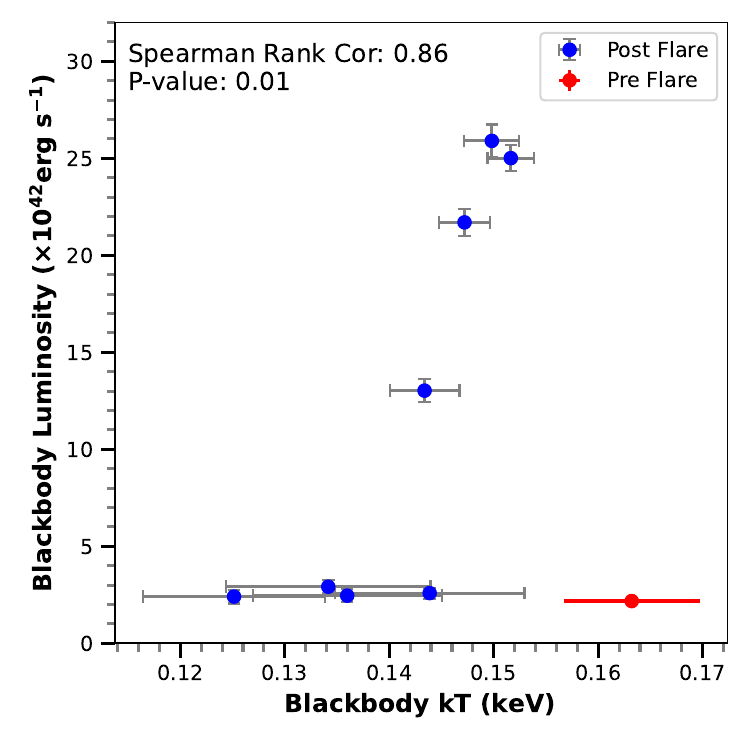}
    \includegraphics[width=7.9cm]{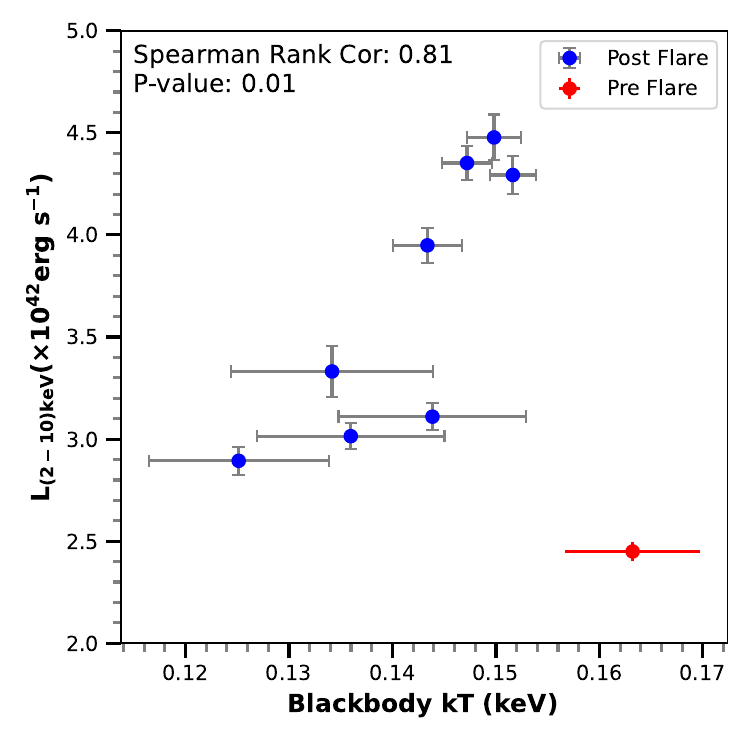}
    }
\hbox{
\includegraphics[width=7.9cm]{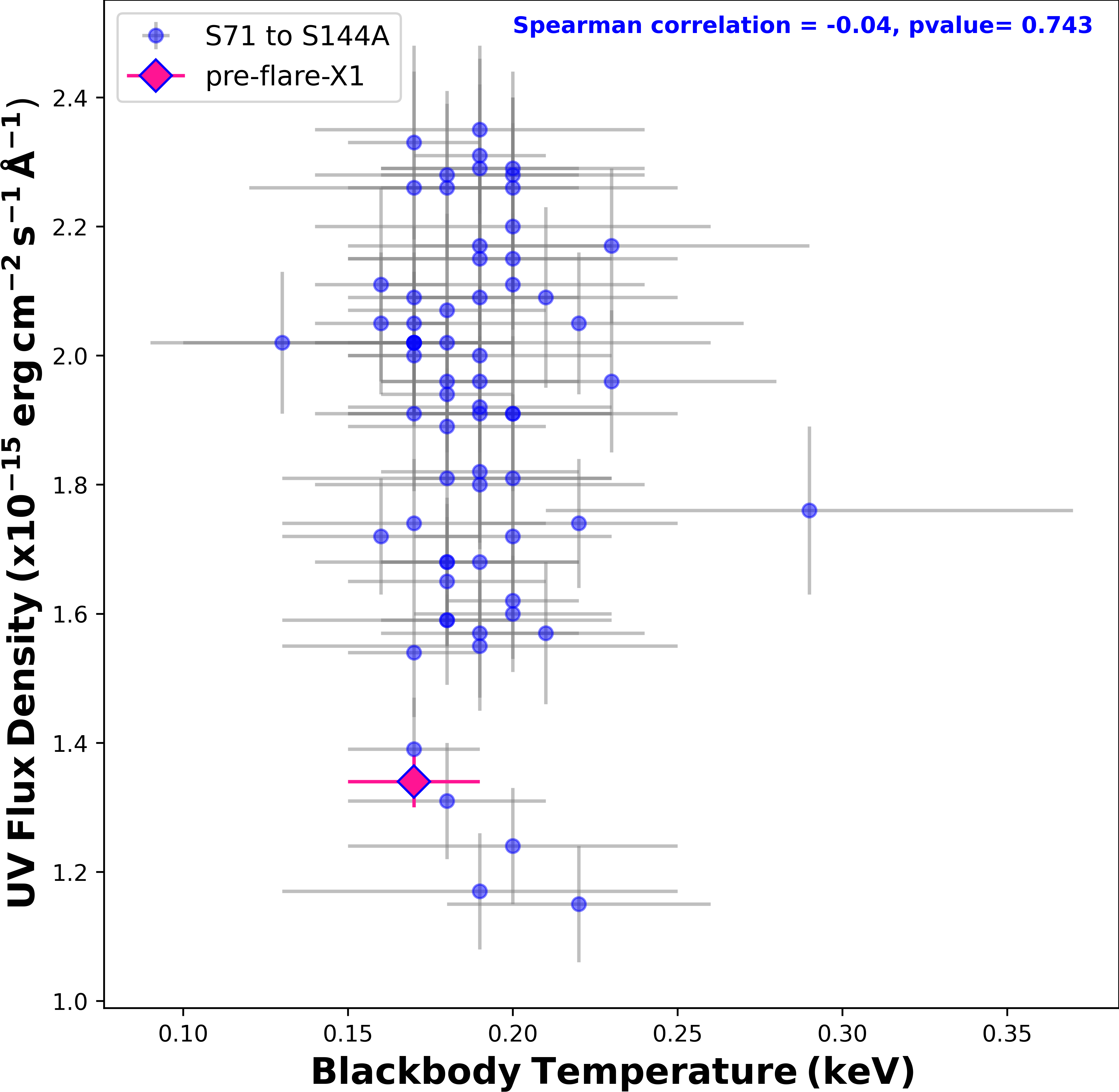}
     \includegraphics[width=7.9cm]{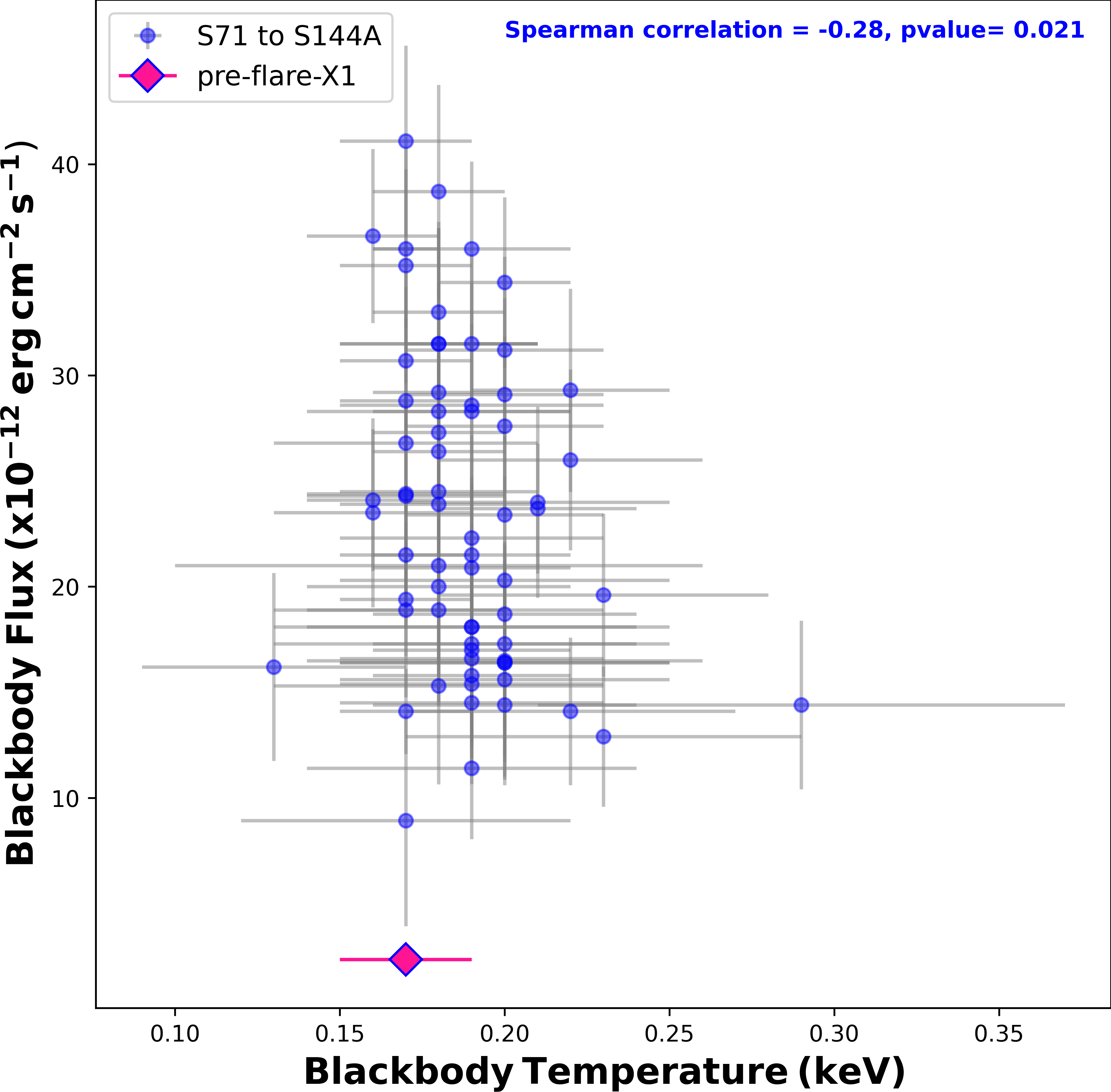}
    }
 }
 \caption{ The correlations between the best fit black body temperature ($kT_{\rm e}$) and the UV flux density, black body luminosity (\xmm{}), black body flux ($\swift$) and power law luminosity (\xmm{}).  {\it Top Left:} The blackbody temperature vs the UVW2 monochromatic flux density {\it Top right:} The blackbody temperature vs power law luminosity (\xmm{}), {\it Bottom left:} The blackbody temperature vs the UV flux density, {\it Bottom right:} The blackbody temperature vs black body flux ($\swift$). We find that the black body temperature is very narrowly constrained and we do not detect any significant correlation between these paramerets.}
 \label{fig:kt_correlations}
\end{figure*}


\bibliography{mybib}
\bibliographystyle{aasjournal}



\end{document}